\documentclass[]{aa}
\usepackage{graphicx}
\usepackage{natbib}  
\bibpunct{(}{)}{;}{a}{}{,}
\usepackage{txfonts}
\usepackage[allcolors=blue]{hyperref}
\usepackage{amsmath}
\usepackage{amstext}

\usepackage{xcolor}
\usepackage{xspace}
\hypersetup{
  colorlinks   = true, 
 urlcolor     = blue, 
 linkcolor    = blue, 
 citecolor   = blue 
}

\newlength\myboxwidth

\defcitealias{Lazorenko2009}{Paper\,I}
\begin{document}

\title{Differential image motion in  astrometric observations with very large seeing-limited telescopes\thanks{\textit{Based on observations made with ESO telescopes at the La Silla Paranal Observatory under programme IDs 086.C-0680, 087.C-0567, 088.C-0679, 089.C-0397, 090.C-0786, 091.C-0083, 092.C-0202, 596.C-0075, and 0103.C-0428.}}}

\author{P. F. Lazorenko\inst{1}
                \and J. Sahlmann\inst{2}
                \and M. Mayor\inst{3}
             	\and E. L. Martin\inst{4}
}
\institute{Main Astronomical Observatory, National Academy of Sciences of the Ukraine, Zabolotnogo 27, 03680 Kyiv, Ukraine                  
                \and   European Space Agency (ESA), European Space Astronomy Centre (ESAC), Camino Bajo del Castillo s/n, 28692,Villanueva de la Ca\~nada, Madrid, Spain 
                		\and
		Observatoire de Gen\`eve, Universit\'e de Gen\`eve, 51 Chemin Des Maillettes, 1290 Versoix, Switzerland  
		\and
   Instituto de Astrof\'isica de Canarias, calle V\'ial L\'actea, San Crist\'obal de La Laguna,
               Spain  
		}
\date{Received ; accepted  }
\abstract{}{{   We investigate how to quantitatively model the observed differential image motion (DIM) in relative astrometric observations. }} 
{{   As a test bed we used differential astrometric observations from the FORS2 camera of the Very Large Telescope (VLT) obtained during 2010--2019 under several  programs of observations of southern brown dwarfs}. The measured image motion was compared to models that decompose atmospheric turbulence in frequency space and translate the vertical turbulence profile into DIM amplitude. This approach accounts for the spatial filtering by the telescope's entrance pupil and the observation parameters (field size,  zenith angle, reference star brightness and distribution, and exposure time), and it aggregates that information into a newly defined {metric integral} term.}
{We demonstrate excellent agreement (within 1\%) between the model parameters derived from the DIM variance and determined by the observations. For a 30~s exposure of a typical 1\arcmin-radius field close to the Galactic plane, image motion limits astrometric precision {to $\sim$60~$\mu$as when sixth-order transformation polynomial is applicable. We} confirm that the measured image motion variance is well described by Kolmogorov-type turbulence with exponent 11/3 dependence on the field size at effective altitudes of 16--18~km, where the best part of the DIM is generated. Extrapolation to observations with extremely large telescopes enables the estimation of the astrometric precision limit for seeing-limited observations of $\sim$5~$\mu$as, which has a variety of exciting scientific applications.}{}

\keywords{Astrometry  --  atmospheric effects -- technique: high angular resolution -- methods: data analysis}
\titlerunning{Differential image motion}
\maketitle

\section{Introduction}{\label{int}}
High-precision {relative} astrometry is a powerful technique used to study the dynamics of various stellar objects, particularly those near the Galactic center (it also has many other astrophysical applications not discussed in this paper). For these purposes, the astrometric precision required should be of 10-100~$\mu$as or better.  For bright objects, this requirement has been well realized with the space mission Gaia \citep{Lind2021}. However, because space telescopes are limited in the size of their light collecting areas, faint objects are measured with  a lower accuracy. 

The new generation of extremely large ground-based telescopes with apertures of about 40 m and adaptive optics have a diffraction-limited resolution, and they can measure very faint stars with a good accuracy. For the Extremely Large Telescope (ELT) \citep{SPIE_ELT, Trippe, ELT2} and the Thirty Meter Telescope \citep{TMT}, the accuracy is expected to be 5 to 40~$\mu$as. However,  ground-based telescopes work in an unfavorable environment (gravity and atmosphere). A common factor is the Earth's atmosphere, which affects the image quality and displaces star positions in the telescope detector, depending on the star color \citep{Pravdo1996,Lazorenko2006}. Another atmospheric effect discussed in this article is the differential image motion (DIM) caused by {high layer} atmospheric turbulence and detected as a random displacement of star images relative to each other. The DIM is closely related to differential atmospheric tilt jitter \citep{Cameron2009}, but it refers to different quantities: the displacement of the image photocenters (not specifying its origin)  and, in the second definition, to the wavefront aberrations, which finally results in the same DIM effect.

The segmented structure of the main mirror, a usual feature of very large telescopes (e.g., ELT, TMT), may affect the image quality. However, when analyzing observations with the OSIRIS camera at the Gran Telescopio Canarias 10.4~m telescope, \citet{GTC} found no  degradation of astrometric precision caused by the segmented primary mirror.

Modern 8-10 m very large telescopes placed in favorable observing conditions also demonstrate rather good astrometric precision. An excellent 100~$\mu$as or better precision was reached on Gemini with a multi-conjugate adaptive optic system \citep{GeMS}. With the Very Large Telescope (VLT) camera FORS2 \citep{FORS} and no adaptive optics, a single-epoch astrometric accuracy of $\sim$100~$\mu$as was reached for 15--20~mag stars  \citep[hereafter \citetalias{Lazorenko2009}]{Lazorenko2009, PALTA1, PALTA2}, { for a series of 30 images}.

Based on a single \citep{Lazorenko2006} or a few series of FORS2 observations \citepalias{Lazorenko2009}, we have already analyzed the DIM effect, but we did not obtain sufficiently reliable results due to the limited data that we used. In this work, we use a large amount of observational data with over ten thousand sky images collected between 2010 and 2019 with FORS2 under observational programs that led to discoveries of a few binary objects \citep{PALTA1, Sahlmann2015DE0823,  Sahlmann_palta3}. The reduction of the observations was based on a specially developed technique \citep{LazLaz} aimed at decreasing DIM. After some modifications, the final version of the method was presented in \citetalias{Lazorenko2009} and \citet{PALTA2}. The approach takes into account the geometric distortion of the field, the motion of the stars in the sky caused by proper motions and parallaxes, the differential chromatic effects dependent on star colors, the zenith distance, the relative motion between CCD chips of the FORS2 detector, and some more minor effects. Also we applied a special procedure for the measurements of star image photocenters in crowded fields, which provided the accuracy at the photon noise limit \citep{Lazorenko2006}. 

{In this article, we try to find a final solution to the atmospheric image motion problem for observations made with very large telescopes that did not use adaptive optics. In spite of the fact that astrometric {precision} is usually limited by the uncertainty of the photocenter measurements, DIM is nevertheless an important component of the error budget, and it has a long history of investigations.}

\section{Outline of investigation}{\label{out}}
The study was run along two independent directions, but each led to the estimate of the same atmospheric DIM variance term  $\sigma^2_\mathrm{a}$. The first line of investigation (Sect.\,\ref{obs}) is purely observational and based on differential reduction of simple imaging observations with the camera FORS2. We derived the residuals $\Delta$  between the measured  and the model-predicted single-frame photocenter position of the target object. {The average square value $\langle \Delta ^2 \rangle$ of the observed differences  $\Delta$ was modeled  by a sum of error components}:
 \begin{equation}
\label{eq:sig_oc0}
 \sigma_\mathrm{mod}^2=\sigma_\mathrm{p}^2 +  \sigma_\mathrm{r}^2 + \sigma_\mathrm{a}^2 + \sigma_\mathrm{extr}^2.   
\end{equation}
Here, $\sigma_\mathrm{p}$,   $\sigma_\mathrm{r}$,  $\sigma_\mathrm{a}$, and  $\sigma_\mathrm{extr}$  are respectively the  uncertainty of the photocenter measurements, the reference frame noise, atmospheric DIM noise, and an additive sum of other unaccounted error terms. The first {two} principal terms (unlike the third term $\sigma_\mathrm{a}$, which depends on the specific atmospheric turbulence above the observation site) are well estimated (Sect.\,\ref{prec}). Thus, we had at hand a well-balanced model of errors. A good visual match between the model $\sigma_\mathrm{mod}^2$ and the observed $\langle \Delta ^2 \rangle$ values for astrometric  observations with FORS2 was demonstrated by some plots in \citetalias{Lazorenko2009}  and \citet{PALTA2}. It means that the bias term $\sigma_\mathrm{extr}$ is small; therefore  $\langle \Delta ^2 \rangle =  \sigma_\mathrm{mod}^2$, and we found the DIM  variance $\sigma^2_\mathrm{a}$ simply by subtracting model variances  $\sigma_\mathrm{p}^2$ and $\sigma_\mathrm{r}^2$ from the observed variance $\langle \Delta ^2 \rangle$. To obtain the best results, we applied a sequence of additional calibrations (Sect.\,\ref{s_cal}) of error components. This step of the investigation is based on the reduction method \citetalias{Lazorenko2009} and \citet{PALTA2} not because of a good astrometric accuracy but because it provides sufficiently reliable estimates of  error terms in Eq.\ref{eq:sig_oc0} and a small component $\sigma_\mathrm{extr}$. 

The second line of investigation is an entirely mathematical description of DIM as a random process analyzed irrespective of actual observations. Spectral consideration of this process (Sect.\,\ref{at}) and use of  data on the measured vertical turbulence  allowed us to find $\sigma_\mathrm{a}$ as a numeric function of parameters related to observations (telescope aperture, exposure,  zenith distance, and sky star distribution) and to astrometric reduction details (order of transformation polynomials $\beta$ used for the reduction to the reference frame). Introduction of the corresponding filter functions simplified the problem to computation of the specific metric integral $I$ (Sect.\,\ref{stand}), which incorporates all observation parameters and gives a solution for $\sigma^2_a$ induced by {each} turbulent layer of the atmosphere. Its final value is a sum over a full atmosphere.

In our final steps (Sect.-s\,\ref{on_int1} -- \ref{s_tr}),  we compared the results of the observational and model lines of investigation and generalized this method for prediction of $\sigma^2_a$ for other very large telescopes, including ELT (Sect.\,\ref{cn2}), that do not use adaptive optics.

\section{Observations and data reduction}{\label{obs}}
{The observational database of the study should ensure extraction of error components in Eq.\,\ref{eq:sig_oc0} with small noise $\sigma_\mathrm{extr}$ and be statistically rich. The observations obtained with the FORS2 camera of the VLT during 2010--2019 for the detection of companions of 20 nearby ultracool dwarfs \citep{PALTA1} meet these requirements well, and }we used images obtained in 14 star fields near the Galactic plane with the best  observational history. Table \ref{ident} provides the list of fields, their sequential numbers (Nr.) assigned in \citet{PALTA1}, the number of stars $N_p$ selected for probing the DIM variance, the average exposure time $T$, the number $M$ of exposures, and $\sec z$ at the average zenith distance of observations. 
 
Each field was observed in a simple imaging mode during 13 to 40 observational epochs of 20--50 exposures per epoch. Observations were made with the $I$ filter, which allowed us to obtain 260 -- 2500 non-saturated field star images per a single exposure. The target was set near the field center. To maintain a good astrometric accuracy, we used a high-resolution collimator, ensuring a  126.1~mas\,px$^{-1}$ CCD scale. 

\begin{table}[tbh]
\caption [] {Target list and summary of observation data. }
{\small
\centering
\begin{tabular}{@{}ccccc|ccccc@{}}
\hline
\hline
Nr.  &$N_p$ &   $T$,\,s & M  & $\sec z$ & Nr. &$N_p$ & $T$, s & M & $\sec z$ \rule{0pt}{11pt}\\
\hline
2   & 2   &  13.1  &  1089 &  1.04 & 14  & 6  & 29.4  & 1120  & 1.21  \rule{0pt}{11pt} \\
6   & 5   &  23.8  &  1202 &  1.05 & 15  & 5  & 28.4  &  558  & 1.09   \\
8   & 6   &  55.7  &   304 &  1.05 & 16  & 4  & 26.0  &  548  & 1.16   \\
9   & 3   &  51.1  &   370 &  1.14 & 17  & 5  & 30.0  &  583  & 1.03   \\
11  & 4   &  50.4  &   322 &  1.16 & 18  & 5  & 36.3  &  458  & 1.03  \\
12  & 5   &  51.8  &   551 &  1.11 & 19  & 6  &  8.5  & 2388  & 1.12 \\
13  & 5   &   3.4  &   842 &  1.15 & 20  & 5  & 28.0  &  540  & 1.12 \\
\hline                      
\end{tabular}
\label{ident}
}
\end{table}

The images were calibrated, and the photocenters ($x_i$, $ y_i$) of all sufficiently bright stars were measured.  Photocenter positions ($x_0$, $ y_0$) of the target measured at exposures $m'=1,2...M$  were transformed to the reference image $m'=0$  using a circular group of reference stars centered at the target. We used  $N_R=11$  discrete field sizes with  $R= 200 \,  (=0.42\arcmin), 280 \ldots 1000$~px (=2.1\arcmin).

Technically, the transformation of star positions from image $m'$ to the reference frame was made as described in \citetalias{Lazorenko2009}  by  the least-squares fit using bivariate functions $f_{iw}(x,y)$  enumerated with a sequential number $w$ and defined by the star $i$ position in frame $m=0$ (reference frame). Explicitly, $f_{i,1}(x,y)=1$,   $f_{i,2}(x,y)=x_i$,  $f_{i,3}(x,y)=y_i$,  $f_{i,4}(x,y)=x_i^2$, \ldots \,   $f_{W}(x,y)= x_i^{\beta_1}y_i^{\beta_2}$, where  $\beta =\beta_1 + \beta_2$  is the highest polynomial order. For each $\beta$, we used a full set of $W=k(k+2)/8$ functions $f_{iw}(x,y)$, with no term omission and assigning effective  weights $P_i=(\sigma^2_a + \sigma^2_\mathrm{p})^{-2}$ for reference stars  $i > 0$ and  $P_0=0$ for the target  $i=0$ \citepalias{Lazorenko2009}. We applied in sequence all orders $\beta$ from 2 to 6 in order to trace dependence of $\sigma^2_\mathrm{a}$  on $\beta$.

In  vector representation,  the best estimates $\boldsymbol{\hat{x}}$  of  positions   $\boldsymbol{x}$ measured in frame $m$ and transformed to the reference frame  $m'=0$  is convenient to express as a linear projection,
\begin{equation}
\label{eq:xy}
\boldsymbol{\hat{x}}=\boldsymbol{ax},
\end{equation}
of the measurements $x$ (similar for $y$), where $\boldsymbol{a= fF^{-1}f^TP}$ is a  $(N+1) \times (N+1)$ matrix of coefficients $a_{i',i}$ with a  projective property $\boldsymbol{af= f}$ and  $\boldsymbol{F}$ is the normal matrix of the transformation. The elements in the first $i=0$ column of $\boldsymbol{a}$ are zero (because we set $P_0 = 0$ for the target), and its first $i'=0$ line contains coefficients $a_{0,i}$ that correspond to  $a_{i}$ { in the model} Eq.\ref{eq:L2}. Hence, the vector of the residuals is
\begin{equation}
\label{eq:dvect}
\boldsymbol{\Delta}=\boldsymbol{x}-\boldsymbol{\hat{x}}.
 \end{equation} 

The uncertainty $\sigma_\mathrm{r}$ of the transformation to the reference frame for a star $i$, either a reference or primary P-star, is called the reference frame noise, and {it is a component term in Eq.\,\ref{eq:sig_oc0}}. According  to \citetalias{Lazorenko2009},  it is equal to  the  ${i,i}$-th diagonal element, 
\begin{equation}
\label{eq:sig_rf00}
\sigma_\mathrm{r}^2 =\boldsymbol{(fF^{-1}f^T)}_{i,i},
\end{equation}
of the covariance matrix. It allows for a double interpretation. In the classic definition, it is the uncertainty of the low-order geometric field distortion correction, and with respect to the spectral description, it is the measure of how accurate low-spectral modes of the DIM are eliminated. In fact, the two interpretations are equivalent.

The full reduction model also includes displacements caused by the proper motion, parallax, and atmospheric chromatic model displacements of stars  \citep{PALTA2} so that  $\Delta$ are free from these effects. In addition, we considered the fact that while astrometric reduction  eliminates geometric field distortion  correspondent to {the $\beta$ order, the next higher-order distortions} remain uncompensated, and {in effect, they work as a term $\sigma_\mathrm{extr}^2$  that biases} $\langle \Delta ^2 \rangle$, and consequently   $\sigma_\mathrm{a}^2$ is shifted to higher values. Fortunately, because within a single series of exposures (time span of 20--40 min) geometric distortions are approximately static, we mitigated this effect by forming differential  $\Delta$ taken relative to its average value for the current series of CCD images, that is by subtracting a single-epoch average of  $\Delta$. This is a critically important step that clears $\Delta$ from static components uncorrelated in time, both of high-order geometric distortion and of the residual imprint of not fully excluded low-order geometric distortion. 

\begin{figure}[tbh]
\includegraphics[width=\linewidth]{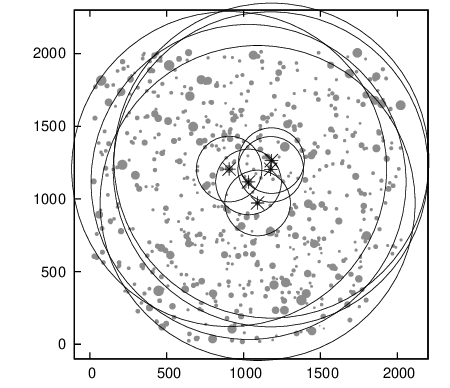}
\caption{All stars (dots) and P-stars (asterisks) imaged near dwarf Nr.12 and the reference fields, shown by small  (for $R=200$~px) and large (for  $R=1000$~px) circles centered at the  corresponding P-star. }
\label{Ftest}
\end{figure}

Astrometric reduction  produced the residuals $\Delta $ for a single  star  (brown dwarf)  in the  field center,  which we then used as the object to probe the DIM variance $\sigma^2_\mathrm{a}$. In this study, we increased the probing star (P-star) number using (in addition to the brown dwarf) some more sufficiently bright stars within  100 -- 200~px of the field center. Stars fainter by more than eightfold of the dwarf were not used because for these stars, the uncertainty of the photocenter measurements (Sect.\,\ref{prec},\ref{s_cal}) is unacceptably large, which  impedes accurate estimates of the DIM effect. Each field was provided with $N_p=2$ -- 6 P-stars  (Table \ref{ident}), including the dwarf itself, thus favoring better estimates of DIM characteristics. An example of P-star distribution used in the field Nr.12 is shown in Fig.\,\ref{Ftest}.  Computations for  additional stars were run independently using their unique set of reference stars considering each P-star in turn as a target. Thus, in each sky field, we derived $N_R \times N_p \times M$ residuals ${\Delta}$.

In addition, we had an opportunity to ingest the measurements of $\Delta$ derived for each P-star when they  were used as reference (except the dwarf itself) for the nearby primary P-star. Despite the fact that {\citetalias{Lazorenko2009} explained that low-order DIM is} completely eliminated for the target only, numerical estimates of  $\sigma^2_\mathrm{a}$ were found to be equal for these two types of objects. In this way we increased the number of single measurements of ${\Delta}$   by a factor of  $(N_p-1)$, improving  statistics of the data.  The number of residuals in a single field is $N_R \times N_p \times (N_p-1) \times M$,  which is of the order of $10^5 $ for each reduction  mode $k$. In total, we collected $12\times 10^6$ residuals.

Finally, we derived the variance $\langle \Delta ^2 \rangle$ of a single image positional residual defined as the weighted average  of ${\Delta}^2$ over $M$ exposures and two coordinate axes. After the rejection of {outlier residuals} (Sect.\,\ref{qq}), we recomputed these estimates.

\section{Error terms in differential measurements and their calibration}{\label{prec}}
For a target specified with $P_i=0$, the {observed variance $\langle \Delta ^2 \rangle$ of the residuals $\Delta$} is modeled by Eq.\ref{eq:sig_oc0}, but for {a subset of reference P-stars} with $P_i > 0$, a valid expression is   $ \sigma_\mathrm{mod}^2= \sigma_\mathrm{p}^2 + \sigma_\mathrm{a}^2  - \sigma_\mathrm{r}^2  $ \citepalias{Lazorenko2009}. For consistency, we united them into a single expression: 
\begin{equation}
\label{eq:sig_oc2b}
 \sigma_\mathrm{mod}^2=\sigma_\mathrm{p}^2 \pm \sigma_\mathrm{r}^2  + \sigma_\mathrm{a}^2,  
\end{equation}
where $\sigma_\mathrm{r}^2 $  is either added for the primary star or subtracted for the reference star. {We omitted the term $\sigma_\mathrm{extr}$ since it was found to be small; thus, we expected that the observed variance $\langle \Delta ^2 \rangle$ would be approximately equal to $  \sigma_\mathrm{mod}^2$. } Furthermore, we admitted a simple approximate relation,
\begin{equation}
\label{eq:sig_pi}
 \sigma_{\mathrm{r},(P_i>0)}^{-2}= \sigma_{\mathrm{r},(P_i=0)}^{-2}+( \sigma_\mathrm{p}^2   + \sigma_\mathrm{a}^2 )^{-1}, 
\end{equation}
between the reference frame noise in the above two cases  \citepalias{Lazorenko2009}.  In the next paragraphs we specify expressions for these components used to find the initial estimates of $\sigma_\mathrm{mod}$, and they are later updated in Sect.\,\ref{s_cal}. 

The component $\sigma_\mathrm{p}$ in Eq.\ref{eq:sig_oc2b} is usually a dominant error component for even the brightest stars, and it is related to the statistical fluctuations of the photon number in pixels of the star image. For bright images and a low background noise, it depends on the full width half maximum (FWHM) and the full electron flux $I_\mathrm{full}$ in the star image. The theoretical expression  $\sigma_\mathrm{p}= FWHM/ (2.35\sqrt{I_\mathrm{full}})$  is given, for example, by \citet{Irwin}, \citet{cramer} for Gaussian star image shape. For FORS2, the Gaussian approximation is not  sufficient; thus, we represented the star profiles by a sum of three Gaussians where the core image component with a sigma parameter $\sigma_\mathrm{G}$ contains a flux $I_G$ approximately equal to two-thirds of the total signal ${I_\mathrm{full}}$  \citepalias{Lazorenko2009}. Two other Gaussians represent the star profile widening in wings. For this compound model, the above-cited expression for $\sigma_\mathrm{p}$ is not valid, and a better empiric expression is
\begin{equation}
\label{eq:sig_ph}
\sigma_\mathrm{p} = c_1 \, \mathrm{FWHM}/(2.34\sqrt{ I_G}),
\end{equation}
where FWHM$=3.10\sigma_\mathrm{G}$ (near $2.35\sigma_\mathrm{G}$, as expected for a Gaussian star profile) and  $c_1=1+0.15(\sigma_\mathrm{G}-1.5)^2$ with  $\sigma_\mathrm{G}$  expressed in pixels. This is a simplified expression valid for bright stars, and a full form is given in \citetalias{Lazorenko2009}. All numeric terms in (\ref{eq:sig_ph}) of course slightly depend on seeing and varying  geometry of star images. Equation\,\ref{eq:sig_ph} is therefore hardly expected to provide an exact estimate of $\sigma_\mathrm{p}^2 $ for any image over a wide 0.5 -- 1.0\arcsec range of the seeing, which is a reason for the few percent of inconsistency between $\langle \Delta ^2 \rangle$ and $\sigma_\mathrm{mod}^2 $.   By applying an extra scale-type calibration  (Sect.\ref{s_cal}) of $\sigma_\mathrm{p}^2 $ (which can refer to numeric terms in expressions for FWHM and $c_1$), we arrived at a sufficiently good match of the model and the measured variances. 

The reference frame noise $\sigma_\mathrm{r}^2 $ for a star $i$, reference or primary, is defined by Eq.\,\ref{eq:sig_rf00}.  It is accurate to within a few percent due to the above-discussed uncertainty in $\sigma_\mathrm{p}$ because it affects elements of the matrix $\boldsymbol{F^{-1}}$. 

As the initial raw model of $ \sigma_\mathrm{a}^2$, we used the empirical expression 
\begin{equation}
\label{eq:sig_at}
\sigma_\mathrm{a}^2 =a (17/T)R^{b}
\end{equation}
{valid for zenith \citep{Lazorenko2006}. Here, the values of $a$ and $b$ are given in Table 1 of that paper  for each  reduction  mode $k$.  To be applicable  in a general case, term $a$ was preliminarily multiplied by a factor $(\sec z) ^{\nu+p+1}$ (Sect.\,\ref{at}).} Because the value of $b$ is between 2.2 and 3.8, the $\sigma_\mathrm{a}^2$ variance quickly increases with $R$.

As commented, the above expressions are approximate only. Therefore, in Sect.\,\ref{s_cal} we update them (individually for each P-star) by means of a calibration based on cross-comparison of the model to the measured variance $\langle \Delta^2 \rangle$. Thus, $\sigma^2_\mathrm{p} $,  $\sigma^2_\mathrm{r}$, and  {   $\sigma_\mathrm{a}^2$  were updated   as explained in Sect.\,\ref{s_cal}, and the following analysis was done with these nominal variances.}

 It is useful to compare the typical values of the terms in the budget of the total noise Eq.\ref{eq:sig_oc2b} for our P-stars. At an average seeing of 0.6\arcsec and 0.1--0.5$\times 10^6$ number of electrons in images,  $ {\sigma}_\mathrm{p} $ is 0.5--1.2~mas. The other two term magnitudes are between 0.1 and 0.5~mas or below, depending on the field star distribution, $\beta$,  exposure, $z$, and $R$. Thus, the uncertainty of the photocenter measurement dominates, except  short exposures,  low $k$ and wide $R$ when $\sigma_\mathrm{a}$ increases.  Also, at narrow $R$ and high $k$ the term $\sigma_\mathrm{r}$ may dominate because of the dependence $\sigma_\mathrm{r} \sim k R^{-1}$ \citep{LazLaz}.  

\section{Model of differential image motion}{\label{at}}
Image motion detected as a random displacement of star photocenter positions is caused by atmospheric turbulence, which introduces light phase fluctuations across the telescope entrance pupil and thus displaces differently the target object and reference star positions. A single layer turbulence concentrated at the altitude $h$ works as the phase screen, which causes differential displacements in astrometric measurements. This random process is characterized by the 2D spectral power density \citep{LazLaz}
\begin{equation}
\label{eq:G}
G(q)= \frac{  0.033(2\pi)^{1-p}C_n^2}{ VT}  Y(q)Q(q)q^{-2-p} \Delta h ,
\end{equation}
where $q$ is a circular spatial frequency related to the screen, $C_n^2$  is the turbulent strength index in this layer,  $V$ is the wind velocity at the screen location,  $\Delta h$  is the layer thickness, and $p$ is a constant, which in the case of the Kolmogorov turbulence is $p= 2/3$.  Because there is some evidence of non-Kolmogorov-type turbulence  \citep{Non_Kolm},  we  treat the term $p$ as a variable until in Sect.\,\ref{s_tr} we find that actually the observed data are fit well with $p= 2/3$, which is expected for Kolmogorov turbulence.  Equation\,\ref{eq:G} is valid for long exposures $T \gg D/V $ and  expresses the process of filtration of the light phase fluctuations by two filters specific to the measurements of differential displacements. The filter $Y(q)= [2J_1(\pi Dq)/(\pi Dq)]^2$  describes averaging of the phase fluctuations across a circular telescope entrance pupil $D$, and the filter $Q(q)$ represents absorption of the DIM power spectrum by the grid of reference stars in differential astrometric measurements. 

While $Y(q)$ effectively reduces the high-frequency DIM  spectrum,  filter $Q(q)$, 
\begin{equation}
\label{eq:L2}
Q(q)= N^{-2} \sum_{i,i'=1}^N a_i a_{i'} [1-2J_0(2\pi d q \rho_i ) + J_0(2\pi d q \rho_{i,i'} )],
\end{equation}
suppresses low-frequency components and represents absorption of the DIM power spectrum by a grid of  $i=1,2 \ldots N$ reference stars imaged in the field of the angular radius $R$. Here, $J_0$ is a Bessel function of the first kind,  $\rho_i $ and $\rho_{i,i'}$ are the angular  distance between the target $i=0$ and the reference star $i'$  or between $i$ and $i'$ reference stars,  $d=h\sec{z}$ is  the inclined distance to the turbulent layer  for observations  made at the zenith distance $z$, and $a_i$ represents values that depend on the astrometric reduction mode $k$  and are specified in  Sect.\,\ref{filt}.   

The integration of function $G(q)$ produces the variance of DIM  
\begin{equation}
\label{eq:sa}
\sigma^2_{\mathrm{a}}= C  I(d)  (T_0/T) \sec{z}
\end{equation}
generated by a single turbulent screen where term
\begin{equation}
\label{eq:ai}
C =    0.033(2\pi)^{2-p}C_n^2   (V T_0)^{-1} \Delta h 
\end{equation}
depends only on the atmospheric turbulence and is numerically  equal to $\sigma^2_{\mathrm{a}}$ at exposure $T_0$ for observations characterized by a unit integral $I$.  The integral
\begin{equation}
\label{eq:int}
I(d)=\int_0^{\infty}  Y(q)Q(q)q^{-1-p} dq 
\end{equation}
is a natural metric of DIM variance in a sense that $\sigma^2_{\mathrm{a}}$ depends on a numeric value of $I$ that inherits most of the principal parameters of the observations: telescope aperture $D$,  the current exposure specification ($R$, $\sec{z}$),  the star distribution in the field, photon flux from stars, background, and the inclined distance $d=h\sec{z}$ to the phase screen. Also, in Eq.\,\ref{eq:sa}, we added a factor $\sec{z}$ to take into account the additional increase of the turbulence proportionally to the depth of the turbulent medium for observations at zenith angle $z$.  Modification of Eq.\,\ref{eq:int} for the von Karman spectrum (finite outer scale of turbulence) is considered in Sect.\,\ref{von}. 

{The variance $\sigma^2_{\mathrm{a}}$ can be reduced for very narrow angle observations when $Rd \ll D/2$,  making the product of filters $Y(q)$ and $Q(q)$ small. } In practice, it forces use of very large ground-based telescopes of the VLT class or bigger to ensure good precision with $R \sim$1--2\arcmin. For  narrow angle observations, the integral $I(d)$ in  Eq.\,\ref{eq:int} permits  approximate dependence, 
\begin{equation}
\label{eq:intappr}
I(d) \sim (Rd)^{\nu+p} D^{-\nu},    
\end{equation}
on $R$, where $\nu$ is a constant related to the entrance pupil shape and which for a circular objective is $\nu=3$. Hence, according to Eq.\,(\ref{eq:sa} - \ref{eq:int}),  the DIM variance $\sigma^2_{\mathrm{a}}$ generated by a single turbulent screen follows the dependence
\begin{equation}
\label{eq:L4}
\sigma^2_{\mathrm{a}}  \sim C  (Rh)^{\nu+p} D^{- \nu}    (\sec{z})^{\nu+p+1} (T_0 / T).
\end{equation}

\subsection{Modal structure of the image motion }{\label{filt}}

Bessel functions in Eq.\,\ref{eq:L2} can be expanded  into an infinite series  of even powers of $q$. Hence, $Q(q)$  has a modal structure  $Q(q)= \sum_{s=1}^{\infty} g_{2s} q^{2s}$ with coefficients $g_{2s}$ dependent on coordinates  $x_i$, $y_i$ of reference stars and the target object  $x_0$, $y_0$  \citep{LazLaz, PALTA2}.  The expressions for coefficients $g_2, g_4..g_{k-2}$ until some even integer mode 
\begin{equation}
\label{eq:beta}
 k = 2(\beta+1 )
 \end{equation}
can be represented \citepalias{Lazorenko2009} as the sum of quadratic terms of the type $[\sum_{i=1}^N a_i f_{iw} - f_{0,w}]^2$. 

Remarkably, all the terms $g_2, ...$ up to and including $g_{k-2}$ can be turned to zero by a proper choice of coefficients $a_i$  that meet  $w=1, \ldots k(k+2)/8$ conditions $  \sum_{i=1}^N a_i f_{iw} - f_{0,w}=0$.   One may note that the least-squares transformation (Sect.\,\ref{obs}) to the reference frame (Eq.\,\ref{eq:xy}) is made with a projection matrix $\boldsymbol{a}$ whose first-line elements meet the above conditions; at the same time, they are used in Eq.\,\ref{eq:L2}. Thus, the least-squares transformation leads to a complete elimination of low expansion modes $2s=2,4 \ldots k-2$  of  $Q(q)$, which is now  nearly opaque for frequencies below about $q_0=k/(4 \pi Rd)$ \citep{LazLaz}.   The result is a cutoff of the  $G(q)$ spectrum at low frequencies, which contains the principal energy of the DIM and consequently leads to a strong decrease of the integral (\ref{eq:int})  and  $\sigma_\mathrm{a}$ values. An even integer $k$, referred hereto as the reduction mode, is the first active (not eliminated) mode of filter $Q(q)$ with dominating energy, so that $Q(q) \approx g_k q^k$.  The effect of the image motion filtration favors use of the high $k$ (or $\beta$) modes because it expands the opaque zone of filter $Q(q)$. 

It makes clear that it is not correct to speak of the DIM variance $\sigma^2_a$ in general as the quantity dependent on the atmosphere only. Instead, we should specify the mode $k$ of the transformation procedure and refer to the DIM variance as a function $\sigma_\mathrm{a}^2(k)$. Thus, the variance  $\sigma_\mathrm{a}^2$ for the reduction with mode $k=6$ is the measure of the DIM spectrum $G(q)$ including the filter $Q(q)$ modes $q^6, q^8\ldots$, while the reduction with $k=8$ is sensitive  to modes $q^8, q^{10}\ldots$. With each subsequent $k$, we  access  different spectral ranges, with a natural decrease of $\sigma_\mathrm{a}^2$ at each next $k$ value. Because the $Q(q)$ filter suppresses frequencies below {$q_0$, we find that the largest linear size of the phase fluctuations related to $Q(q)$ is 
\begin{equation}
\label{eq:Lcut}
L_Q  \simeq 4 \pi Rd /k.
 \end{equation}
For instance, at a distance of $d=18$~km to the turbulent layer, a field size $R=2.1\arcmin$, and  $k=6$,  its value is $L_Q \simeq  20~m $ irrespective of $D$. In particular, this means that the DIM is only marginally affected by the actual value of the outer scale of turbulence if it is longer than 20~m for telescopes of VLT or ELT size (Sect.\,\ref{von}).}

\subsection{Metric integral $I$ and its properties}{\label{stand}}
The measured DIM variances  (estimates of $\sigma^2_\mathrm{a}$) are not homogeneous and not directly comparable between fields because they correspond to different $z$. Of course, it is easy to reduce ${\sigma}^2_\mathrm{a} $ to zenith using the approximate dependence $\sigma^2_\mathrm{a} \sim (\sec{z})^{\nu+p+1}$  of DIM variance on $z$ valid for narrow fields   (Eq.\ref{eq:L4}), but it is possible to apply exact conversion, leading to a better accuracy. Another problem arises when we want to develop a universal model for $\sigma^2_\mathrm{a}$ that allows us to predict its value for any star field and for any altitude of the telescope above the sea level. In this case, we should take into consideration the fact that the  ${\sigma}^2_\mathrm{a} $ value depends on the distribution of the stars. Moreover, they are not comparable even for P-stars within the same field because of essentially different values of $I$ defined by Eq.\,\ref{eq:int}. All of these problems are solved with the introduction of a standard star distribution. As this standard template, we considered easily handled continuous distribution of reference stars (CSD). The properties of $I$ for the standard distribution allowed us to model $\sigma^2_\mathrm{a}$ based on the atmosphere turbulence data (Sect.\ref{on_int}) with a subsequent comparison to FORS2 observations.

\subsubsection{Continuous  distribution of reference stars}{\label{demoint}}
The value of $\sigma^2_\mathrm{a}$  is a linear function Eq.\,\ref{eq:sa} of the metric $I(d)$, which via the filter $Q(q)$ significantly depends on the star distribution in the field and the brightness of the stars, and therefore, it is strongly variable. 
Its properties are easily analyzed in the tutorial case of continuous (non-discrete) distributions of equally bright stars completely filling the circular reference field and placing the target object exactly in the field center. These distributions serve as a good model of the FORS2 field's properties, which in this study are located near the Galactic plane and are very rich in stars. We consider them to be a template distribution with expressions for  $Q(q)$ available as functions \citep{LazLaz}
\begin{equation}
\label{eq:6_7}
\begin{array}{@{}l@{}l@{}l@{}}
  &  \left [ 1 + \frac{4J_1(y)}{y}- \frac{24J_2(y)}{y^2} \right ] ^2,  & k=6,8;\, \beta=2,3 \\
Q(q)= & \left  [ 1 - \frac{6J_1(y)}{y} 	+\frac{96J_2(y)}{y^2}- 	\frac{480J_3(y)}{y^3}  \right ] ^2, 			& \,k=10,12; \beta=4,5 \\
 & \left  [ 1 + \frac{8J_1(y)}{y} 	-\frac{240J_2(y)}{y^2}+ 	\frac{2880J_3(y)}{y^3} \right. & \\    
 & \left. -\frac{13440J_4(y)}{y^4}  \right ] ^2, 			& k=14,16; \,\beta=6,7 \\
\end{array}
\end{equation}
of the variable $y=2\pi  d q R$, where $dR$ is the linear size of a star field at distance $d$. Above, we added one more expression for $k=14$, 16 derived using the analytic approach in Sect.\,6.1 of the paper cited, where a factor $(k/4-i+1)!$ in Eq.\,41 is (due to a misprint) to be read as $(k/4+i-1)!$. For  odd $k/2$ (for $k=6$, 10, 14 or $\beta=2$, 4, 6)  functions, the value of $Q(q)$ is identical to the nearby next even modes $k/2$ (for $k=8$, 12, 16  or $\beta=3$, 5, 7) because the least-squares solutions are identical in these cases at the target location (of course, solutions in other field points are different, but we note that function $Q$ is defined for the field center only). Functions $Q(q)$ in Eq.\,\ref{eq:6_7} are expanded into powers of spatial frequencies with the first non-zero member $q^k$,  with all  lower expansion terms turned to zero (Sect.\ref{filt}). The integration Eq.\,\ref{eq:int} that uses expressions from Eq.\,\ref{eq:6_7}  yields  $I_\mathrm{c}(d,k,R)$ values  for CSD as a function of  $d$, $R$, and $k$, but in the following, we stick to a shorter designation $I_\mathrm{c}(d)$. Otherwise, we explicitly specify the variables. 

\begin{figure}[tbh]
\includegraphics[width=\linewidth]{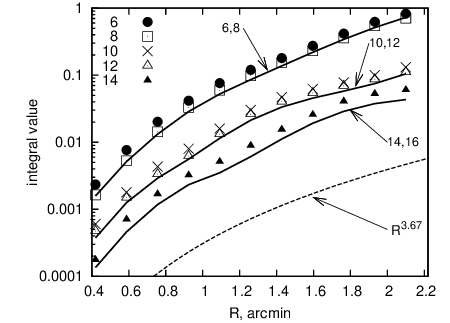}
\caption{Metric integral $I$ dependence on the  field size $R$ for CSD (thick solid lines) and for  FORS2 fields in average (different symbols). Approximate dependence $I \sim R^{3.67}$ (dashed line). All integral values were computed at $d=18$~km.}
\label{tr}
\end{figure}

Figure\,\ref{tr} makes a general impression on the absolute magnitude of $I$ for CDS and for FORS2 fields  for some parameters $R$ and $k$. Both integrals approximately follow a power dependence $R^{\nu +p}$; therefore, their change is over three decades within a quite moderate range of $R$.

\begin{figure}[tbh]
\includegraphics[width=\linewidth]{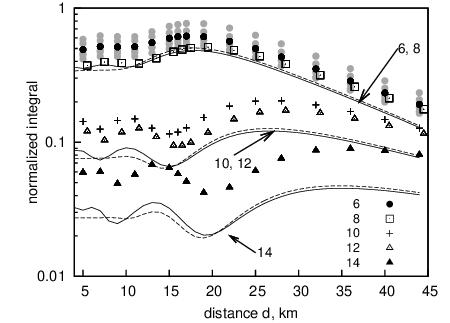}
\caption{ Normalized integral  ${I}_\mathrm{c}^*(d)$ dependence on distance $d$ to the turbulent phase screen for CSD if the  VLT design is with (solid lines) or without a secondary mirror (dashed lines). Integral  $\langle I_\mathrm{F}^*(d)\rangle$  averaged over all P-stars (different type symbols for each $k$ mode shown) within a sky field No.12.  Integral  ${I}_{\mathrm{F},i}^*$  for each P-star in those fields (gray symbols, $k=6$ only). Integrals were computed for $R=2.1\arcmin$} 
\label{fgint}
\end{figure}

At some fixed $R$, the  values of $I_\mathrm{c}(d)$ increase quickly with $d$ following a power law $I_\mathrm{c}(d,R) \sim d^{\nu+p}$ (\ref{eq:intappr}) valid for narrow fields but not exactly. In order to examine minor deviations from the power law, we formed integrals ${I}_\mathrm{c}^*(d)= I_\mathrm{c}(d)(d_0/d)^{\nu+p}$  normalized to some reference distance $d_0$ and computed at some fixed $R$. We adopted  $d_0= 16$~km  because  for this distance, we initially computed integrals $I_{\mathrm{F},i}(d_0)$  for each $i$-th P-star in the FORS2 fields. For illustration,  Fig.\,\ref{fgint} presents normalized functions ${I}_\mathrm{c}^*(d)$  versus  distance $d$ to the phase screen. These functions vary quite moderately within a full range of distances, while the original function $I_\mathrm{c}$ change is $\sim 1000$. 

We present $I_\mathrm{c}^*$ functions computed in two ways: one for a fully open circular entrance aperture of diameter $D=8.2$~m and another for the actual VLT optical design with a secondary mirror of $D_\mathrm{s}= 0.9$~m diameter. In the latter case, instead of Eq.\,\ref{eq:G}, we used a modified  expression  $Y(q)=[ 2 J_1(\pi Dq)/(\pi Dq) - (D_\mathrm{s}/D)^2   2 J_1(\pi D_\mathrm{s}q)/(\pi D_\mathrm{s}q) ]^2$  derived from a difference of the Fourier transforms of two mirrors.  The computed integrals in both cases are quite similar in shape, but in the following, we keep to the version with a secondary mirror.

The original metric integrals  $I_\mathrm{c}$ and their modifications $I_\mathrm{c}^*$  in Fig.\,\ref{fgint} were computed for even $k/2$ only because functions $Q(q)$ and, consequently, the integrals for nearby smaller $k$ are the same. For instance,  the integral values for $k=6$ and $k=8$ are equal. Therefore, use of the higher $k=8$ mode instead of $k=6$ results in no decrease of ${I}_\mathrm{c}^*$,  but the next $k=10$ mode leads to a significant, roughly fivefold, decrease of ${I}_\mathrm{c}^*$ (and consequently of the DIM variance $\sigma^2_\mathrm{a}$).  
                     
We also note that Fig.\,\ref{fgint} refers to $R=2.1\arcmin$. For other $R'$, in accordance with the definition of the $y$ variable in Eq.\,\ref{eq:6_7}, the $I_\mathrm{c}^*(d)$ shape is the same if plotted versus a distance scaled with a factor $R'/R $. 

\subsubsection{Metric Integral $I$  for FORS2 fields}{\label{demointF}}
The integral $I$ values for the FORS2 fields are easily evaluated in terms of effective reference field size $S= \left | N^{-2} \sum_{ii'}^N a_i a_i' d^k (\rho_i^k+\rho_{i'}^k-\rho^k_{ii'}) \right |^{1/k}$; however, these estimates are approximate only \citep{LazLaz}. In this study, we wanted to meet the best accuracy provided by diverse observational statistics. Therefore, we computed $I$ values by direct integration. The numerical results were found to be quite stable to seeing conditions, and therefore, the  computations were done using a single typical image in a sky field. We revealed quite similar features of this integral dependence on $d$ for FORS2  and for CSD.  This is illustrated, for example, in the fields No.\,12  and 14, for which we computed  $I_{\mathrm{F},i}$ for each calibration star $i$. For these fields, the normalized functions  ${I}_{\mathrm{F},i}^*(d)  =I_{\mathrm{F},i}(d)(d_0/d)^{\nu+p}$ are quite similar in shape to the functions $I_\mathrm{c}^*(d)$ of the corresponding order $k$, but they differ a little in magnitude (Fig.\,\ref{fgint}). A useful quantity specifying this feature is the ratio  
\begin{equation}
\label{eq:mu}
 \mu_{i} = I_{\mathrm{F},i}^*(d)/I_\mathrm{c}^*(d) =I_{\mathrm{F},i}(d)/I_\mathrm{c}(d) 
 \end{equation}
 formed for each $i$-th calibration star. We find that it nearly does not depend on $d$, so we could assume that  $\mu_{i} = $ const within a sky field for each $i$ at a fixed $k$ and $R$. At the same time, in different fields, its magnitude can vary between 0.5 and 1.5, with a scatter more often detected for $k/2$ odd ($k=6$, 10, 12).  Especially distinct variability is detected at smaller $R \approx 0.5\arcmin$. Even in a single field variation of $\mu_i$ for the nearby calibration stars are quite significant (compare the scatter of individual $I_{\mathrm{F},i}$ marked by gray symbols in Fig.\,\ref{fgint}) but natural and caused by use or exclusion of only a single bright (nearby calibration) star in the list of reference stars. A stronger concentration of bright stars in the field center leads to smaller $I_{\mathrm{F},i}$ values and the least amount of scatter of their individual values.

\begin{figure}[tbh]
\includegraphics[width=\linewidth]{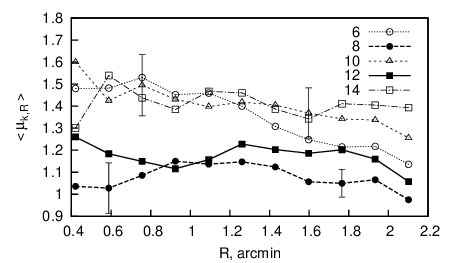}
\caption{Average excesses $\langle\mu_{k,R} \rangle$  of integrals $I_{\mathrm{F}}$  for FORS2 over $I_{\mathrm{c}}$ for CSD, for each reduction mode $k$ (different symbols). Approximate 1-sigma uncertainty of $\langle\mu_{k,R} \rangle$ due to variations  of $\mu_{k,R}$ in separate sky  fields  (error bars). }
\label{mu}
\end{figure}

The significant variability of $I_{\mathrm{F},i}$ in star fields should be taken into consideration in this study in cases (Sect.\ref{s_tr}) where we want to derive image motion characteristics for a typical (average) FORS2 field. Therefore, to remove the variable component in $\mu_{i}$ dependent on the sample, we averaged these values over calibration stars and star fields and obtained the terms  $\langle\mu_{k,R} \rangle$ or typical excesses of $I_{\mathrm{F},i}$ over $I_\mathrm{c}$ at fixed $k$ and $R$, shown in Fig.\,\ref{mu} as a function of $R$. We note that we found a small value for the excess of $\mu$ over a unit for even $k/2$, which indicates a good astrometric quality of  reference fields near the Galactic plane. For odd $k/2$,  the discreteness of the star distributions led to a large,  about 30\%,  excess at $R=2.1$\arcmin. For narrower fields, this effect increases further to  $\sim50$\%. Thus, for odd $k/2$, use of the next $k+2$ reduction mode decreases the  variance $\sigma^2_\mathrm{a}$ by the values indicated. 
 
Some useful expressions for the computation of the FORS2 metric integral at any $d$ and $R$  are given in Sect.\,\ref{int22}. In the following analysis, we often refer to these expressions.

\subsection{Model of the DIM variance based on the turbulence profile }{\label{on_int}}

\begin{figure}[tbh]
\includegraphics[width=\linewidth]{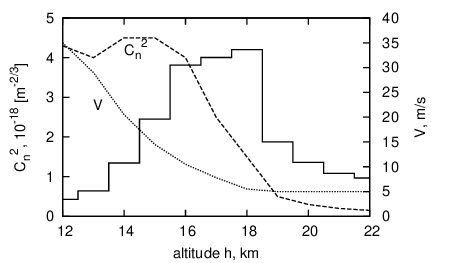}
\caption{Relative input to the DIM variance  (steps) induced by turbulent layers with the measured turbulent index $C_n^2$ (dashed) and the  wind velocity $V$ (dotted). Altitude $h$ is above the VLT telescope.}
\label{fgcn}
\end{figure}

As commented in Sect.\ref{at},  a single atmospheric  turbulent layer  causes the DIM variance $\sigma^2_{\mathrm{a}}$ expressed by Eq.\,\ref{eq:sa}. A full  cumulative atmosphere effect in the zenith is a sum over all layers
 \begin{equation}
\label{eq:turb0}
\sigma^2_\mathrm{a}= (T_0/T) \sum_j  C_j I (h_j),
\end{equation}
where $h_j$ is the layer altitude above the telescope (2.65~km above the sea level) and $C_j$ {(defined by Eq.\,\ref{eq:ai})} is its input to the DIM variance at exposure $T=T_0$ and $I (h_j)=1$. The impact of the layers strongly depends on $h_j$   because   $I(h_j) \sim h_j^{\nu+p}$, in accordance with Eq.\,\ref{eq:L4}.   This is illustrated by Fig.\,\ref{fgcn}, where $C_n^2$ is the median turbulent index for the VLT site, which we adopted as a combination of observed data \citep{Osborn2018, SCIDAR1, SCIDAR2},  and the wind velocity  provided by radio soundings  \citep{MOSE}. The range of the effective altitudes is narrow and limited by 16 -- 18~km only because the DIM is strongly suppressed at low altitudes by the factor  $ h^{\nu+p}$ peculiar to astrometric reduction procedure and, at higher $h$, by the decrease of $C_n^2$. Within this narrow range of $h$, approximation (\ref{eq:Ii_d}) is precise enough to estimate $I(h_j)$ at any $h_j$ via its value at some reference altitude $16 \leq h_\mathrm{ref} \leq 18$~km as  $I(h_j)= F(h_j)/F(h_\mathrm{ref}) I(h_\mathrm{ref}) (h_j/h_\mathrm{ref})^{\nu+p} $ because a ratio $F(h_j)/F(h_\mathrm{ref})$ is  nearly  a unit within this range of altitudes. We let $h_\mathrm{ref}=18$~km, and hence we arrived at a rough but useful approximation, 
\begin{equation}
\label{eq:sigmod}
\sigma_0^2  =A_0 I(h_\mathrm{ref}) (T_0/T), 
\end{equation}
for $\sigma^2_\mathrm{a}$,  where
\begin{equation}
\label{eq:turb}
A_0 = \tau_0 \sum_j  C_j (h_j/h_\mathrm{ref})^{\nu+p}.
\end{equation}
The above expressions allowed us to easily find $\sigma_0^2 $  by summation over turbulent layers, which is the estimate of $\sigma^2_\mathrm{a}$ at zenith. In view of the following results, we inserted a factor  $\tau_0=0.84$ just to make roughly equal the model and the DIM variances measured with FORS2. With the measured turbulence data adopted and shown in Fig.\,\ref{fgcn} and $T_0=30$~s, we found  $A_0= 44.5~$mpx$^2=   0.708$~mas${^2}$. It is equal to $\sigma^2_0$ in the case of an imaginary star field with a unit integral $I({h_\mathrm{ref}})$ and $T=30$~s; thus, $\sigma_0=0.841$~mas.  Half of the variance $\sigma^2_0$, or 22.3~mpx$^2=   0.35$~mas${^2}$, is generated at 16 -- 18~km altitudes. For  fields with an arbitrary star distribution, $\sigma^2_0$ is consequently computed with Eq.\,\ref{eq:sigmod}, which by its content coincides with Eq.\,(\ref{eq:sa}) for a single layer at $z=0$ but generalized to the full atmosphere; in addition, it contains the predicted amplitude $A_0$. The integral $I(h_\mathrm{ref})$ depends on $R$ and the reduction mode $k$ and should be computed using Eq.\,\ref{eq:int} with the $Q(q)$ function with the actual  star distribution and the corresponding brightness at  $d=18$~km. 

The estimate of $A_0$ refers to some average atmospheric conditions, but we note that instant  turbulent profiles are strongly variable \citep[Fig.~2]{MCAO}. For instance, the rms variability of $C_n^2$ and $V$ per day mounts roughly to 50\% of its average value  \citep{MOSE}; therefore, a similar uncertainty is expected in the DIM estimates based on Eq.\,\ref{eq:turb}. Some deviations between the sources of the turbulence profile measurements we used and a lack of precise data on $V$ at high altitudes lead to a roughly 20\%  uncertainty in the prediction of  $\sigma^2_\mathrm{a}$; therefore, we expect that our estimate of $A_0$ is accordingly approximately accurate. In particular, not having a unit value of $\tau_0$ is probably due to an inconsistency between the adopted profile in Fig.\,\ref{fgcn} and the actual turbulent profiles, {or, alternatively, due to the finite outer scale of turbulence (Sect.\,\ref{vonM})}. The actual variations of $C_n^2$ for a single night mount to an even larger $\pm 50\%$ relative uncertainty of the DIM. 

The expressions (\ref{eq:sigmod}--\ref{eq:turb}) are applicable to any $R$ and $k$, as they are universal in this  respect, and they are easy to apply. However, they were obtained assuming the exact validity of the dependence $I(h_j)= I(h_\mathrm{ref})(h_j/h_\mathrm{ref})^{\nu+p} $. More accurate estimates  based on this  model can be obtained using numeric values of integrals $I_{\mathrm{c}}(d)$ or $I_{\mathrm{F}}(d)$ expressed by Eq.\,\ref{eq:Ii_Ic} for summation over altitudes, which results in the variances 
\begin{equation}
\label{eq:calc}
\begin{array}{l}
\sigma^2_{\mathrm{c}}(k,R) = (T_0/T)  \sum_j  C_j  I_\mathrm{c}(h_j,k,R)   \\
\sigma^2_{\mathrm{F},i}(k,R)=  (T_0/T)  \mu_{i,k,R} \sum_j  C_j   I_\mathrm{c}(h_j,k,R)
\end{array}
\end{equation}
representing estimates of  $\sigma^2_\mathrm{a}$  for each $R$ and $k$. Here, the second line refers to the DIM variance {   in FORS2 fields}  for an individual star $i$, and it contains the factor $\mu_{i,k,R}I_\mathrm{c}(h_j,k,R)$ defined by  Eq.\ref{eq:mu} and which in the narrow range of effective altitudes is nearly independent on $h_j$. It is useful to rewrite Eq.\,\ref{eq:calc} as 
\begin{equation}
\label{eq:cf}
\begin{array}{l}
\sigma^2_{\mathrm{c}}(k,R)= A_{\mathrm{c}}(k)   I_\mathrm{c}(h_\mathrm{ref},k,R) (T_0/T) \\
\sigma^2_{\mathrm{F},i}(k,R)=A_{\mathrm{c}}(k)  I_{\mathrm{F},i}(h_\mathrm{ref},k,R) (T_0/T)\\
\end{array},
\end{equation}
 keeping a linear dependence of the variance on  $ I_\mathrm{c}(h_\mathrm{ref})$ or $ I_{\mathrm{F},i}(h_\mathrm{ref})$  (computed at a given $k$ and $R$) just as in the approximate model (\ref{eq:sigmod}), but instead of $A_0$, there are a better defined terms
\begin{equation}
\label{eq:ack}
A_{\mathrm{c}}(k) =   \sum_j  C_j  I_\mathrm{c}(h_j,k,R)/ I_\mathrm{c}(h_\mathrm{ref},k,R)  
\end{equation}
 that are functions of $k$ and $R$ and  universally defined both for the CSD and  FORS2 fields. We computed $A_{\mathrm{c}}(k)$ for each $R$ and found that the variation of these terms with $R$ is within $\pm 10\%$, and it can therefore be neglected. Thus, in Table~\ref{acf}  we give their median over  $R$ as an adequate approximation. Also, because the variation of values $A_{\mathrm{c}}(k)$ with $k$ is  moderately low, we find that sometimes it is reasonable to represent them roughly with a single quantity $A_{\mathrm{c}}(\langle k \rangle)= 0.701$~mas$^2$ (or  $\sigma_{\mathrm{c}}= 0.837$~mas) that is the weighted average over $k$.  We conclude that Eq.\,\ref{eq:cf}, which is  the  updated version of the model (\ref{eq:sigmod}), leads to the estimate close to the amplitude $A_0= 0.708$~mas$^2$ obtained for a simpler model based on approximation of the very narrow fields.

Furthermore, we found that the variances  $\sigma^2_\mathrm{c}$ and $\sigma^2_\mathrm{F}$ computed with   Eq.\,\ref{eq:calc} are numerically well fit by a power dependence 
 \begin{equation}
\label{eq:Ac}
\begin{array}{l}
\hat{\sigma}^2_{\mathrm{c}}(k,R)  =  a_{\mathrm{c}}^2(k)F^2_{\mathrm{eff}}(R) (R/R_0)^{\nu +p} (T_0/T)  \\ 
\hat{\sigma}^2_{\mathrm{F}}(k,R)  =  a_{\mathrm{F}}^2(k)F^2_{\mathrm{eff}}(R) (R/R_0)^{\nu +p}(T_0/T).  \rule{0pt}{11pt} \\
\end{array}
\end{equation}
on the field size $R$. This representation  follows  from   Eq.\,(\ref{eq:I_r}), and $F^2_{\mathrm{eff}}(R)$ is the  effective value of  the function $F(d_0R/R_0)$  accumulated over all layers. The fit terms $a_{\mathrm{F}}^2(k)$ we computed in approximation $F^2_{\mathrm{eff}}(R)=1$ and by replacing $I_{\mathrm{F},i}(h_\mathrm{ref})$ in Eq.\,(\ref{eq:cf}) by  $\langle\mu_{k,R} \rangle  I_\mathrm{c}(h_\mathrm{ref})  $ therefore $\sigma^2_{\mathrm{F}}(k,R)$ refer to typical FORS2 fields.   The amplitude terms computed  for $R_0=1\arcmin$  and $T=T_0$ are given in Table~\ref{acf}. We find that  $a_{\mathrm{F}}^2(k) > a_{\mathrm{c}}^2(k) $, mostly for $k=6$, 10,  14, which is in accordance with the excess in function $\langle \mu \rangle$ values demonstrated by Fig.\ref{mu}.

{We summarize this section with a remark that the $\sigma^2_{\mathrm{a}}$ can be modeled either by direct summing Eq.\,(\ref{eq:I_r}) over each layer input to the DIM variance or alternatively by compressing all the turbulence into a single layer at $h_\mathrm{ref}$, with an amplitude Eq.\,(\ref{eq:ack})  caused by all the atmosphere (Eq.\,(\ref{eq:cf})). } In the following, we compare the derived estimates based on the turbulence profile and on the actual  observations.

\begin{table}[tbh]
\caption [] {Parameters  $A_{\mathrm{c}}(k)$,   $a_{\mathrm{c}}(k)$  for  CSD   and $a_{\mathrm{F}}(k)$  for  FORS2 fields  based on the turbulence profile at a 30~s exposure.}
{\small
\centering
\begin{tabular}{@{}c|@{  \, }c@{  \, }c@{  \, }c@{  \, }c@{  \, }c@{  \, }c@{}}
\hline
\hline
 parameter          &\multicolumn{6}{c}{  $k$ / $\beta$}     \rule{0pt}{11pt}\\
\cline{2-7}
              &      6 / 2        &    8 / 3         & 10 / 4      & 12 / 5       &  14 / 6    & average \rule{0pt}{11pt}\\
\hline                                     
$A_{\mathrm{c}}(k)$, mas${^2}$ &   0.690      &   0.690      &   0.754         & 0.754        &  0.774    & 0.701  \rule{0pt}{11pt} \\
$ a_{\mathrm{c}}(k)$ $\mu$as   & 181$\pm 1$ &  181$\pm 1$ &   76$\pm 1$  & 76$\pm 1$ &   45$\pm 1$     & -  \\
 $a_{\mathrm{F}}(k)$ $\mu$as  & 204$\pm 4$  &   185$\pm 4$ & 90$\pm 2$ &  82$\pm 1$ &  55$\pm 1$  & -\\
 \hline                                     
\end{tabular}           
\label{acf}
}
\end{table}

\section{DIM variance derived from FORS2 observations}{\label{on_intF}}

According to  Eqs.\,\ref{eq:sigmod},\,\ref{eq:cf}, we expected that the DIM variance for FORS2 observations, in zenith, to be proportional to  $I(h_\mathrm{ref})$. This is valid if our concept of the DIM effect (Sect.\,\ref{at}) is correct and, on the other side, astrometric reduction of FORS2 observations and extraction of the DIM variance were performed sufficiently good. This was a key item of investigation to verify in the next Sections.

\subsection{Conversion to zenith}{\label{convZ}}
At first, we reduced {the observed variances ${\sigma}^2_\mathrm{a} $}, which refer to some zenith angle $z$, to zenith. As discussed in Sect.\ref{on_int}, the effective range of the altitude of the turbulent layers is very narrow, from 16 to 18~km. For this reason, we could assume that the DIM effect from a full atmospheric depth is approximately equal to that from a  single layer located at some reference altitude $h_\mathrm{ref}$ within this range. Then, according to  Eq.\,\ref{eq:sa},  the measured DIM variance is proportional to  $I(h_\mathrm{ref} \sec{z})$. Using conversion (\ref{eq:Ii_d}) of the integral from the inclined distance $h_\mathrm{ref}\sec{z}$ to  $h_\mathrm{ref}$,  we found that the measured estimate of $\sigma^2_\mathrm{a}$ at zenith is
\begin{equation}
\label{eq:snorm0}  
{\sigma}^2_{\mathrm{a}}\left|_{z=0}(k,R)={\sigma}^2_{\mathrm{a}}\right|_{z>0}[F(h_\mathrm{ref}/F(h_\mathrm{ref} \sec{z})](\sec{z})^{-\nu-p-1}.
\end{equation}
In this way, we transformed the observed variances to conditions comparable for all fields, thus arriving at a uniform dataset to be analyzed jointly. Conversion (\ref{eq:snorm0}) is quite stable in respect to $h_\mathrm{ref}$ choice because the ratio $F(h_\mathrm{ref}/F(h_\mathrm{ref} \sec{z})$ is  not very sensitive to this term value. Further reduction to the standard exposure $T_0=30$~sec provided variances
\begin{equation}
\label{eq:snorm}
{\sigma}^2_{\mathrm{a}}(k,R)=\left.\sigma^2_{\mathrm{a}}\right|_{z=0}(T/T_0),
\end{equation}
 that are also not dependent on $T$.

\subsection{Dependence of FORS2 DIM variance on the integral $I$}{\label{on_int1}}

Considering that the location of the effective turbulence is likely within 16 and 20~km, reduction to zenith was made for a sequence of reference altitudes $h_\mathrm{ref}$ in this  range,  with  the results, in terms of $\hat{\sigma}^2_{\mathrm{F},i}$, deviating within $\pm 2\%$ only. In parallel, we converted $I_{\mathrm{F},i}(d_0)$, computed originally for  $d_0=16$~km, to this set of distances $d=h_\mathrm{ref}$ using  Eq.\,\ref{eq:Ii_d}. In this way we converted both the observed DIM  variances and the integrals $I_{\mathrm{F},i}$ to  conditions uniform for all fields irrespective of $z$ and $T$. 

We proceeded with validation of the approximate model Eq.\,\ref{eq:sigmod}, assuming that $\sigma_0^2$ and $ I(h_\mathrm{ref})$ in this equation are represented, respectively, by a set of  the measured values  ${\sigma}^2_{\mathrm{a},i}(k,R)$ and $ I_{\mathrm{F},{i}}(h_\mathrm{ref},k,R)$ for each $i$-th P-star irrespective of its status (primary or reference).  Thus, we wanted to find whether ${\sigma}^2_{\mathrm{a},i}(k,R)$ is a linear function of $ I_{\mathrm{F},{i}}(h_\mathrm{ref})$  not taking dependence on the $k$ mode into consideration. The immediate fit of these data with  Eq.\,\ref{eq:sigmod}, however, revealed a certain systematic pattern in the fit residuals at small $R\leq 1\arcmin$ very similar in all sky fields for each separate $k$. We found that this inconsistency is well modeled by two offset terms, $\mathrm{off}_k$ and $\delta^*_k$, that are introduced for each $k$. The first term is a general offset to the DIM variance, and the second is a correction to the zero-point of the adopted reference frame uncertainty (parameter $\delta_r$ in Eq.\,\ref{eq:m1}). These terms were incorporated  for each $k$ mode separately, and instead of  Eq.\,\ref{eq:sigmod}, we used the updated model of DIM variance,
 \begin{equation}
\label{eq:sa2}
{\sigma}^2_{\mathrm{a},i}(k,R)  =  \hat{A}_0  I_{\mathrm{F},{i}}(h_\mathrm{ref}) + \mathrm{off}_k + \delta^*_k R^{-2}_i,
\end{equation}
to fit the full set of  measurements combined for all  P-stars, $k$, and $R$. Here, $\delta^*_k R^{-2}_i$ is a minor approximate correction to $\sigma_{\mathrm{r},i}^2$. Both offsets produce an effect below  $\pm 0.005$~mas$^2$ (= $\pm 0.3$~mpx$^2$), or about 2 -- 4\% of the DIM variance magnitude in the range of $R $ from 0.7 to 2.1\arcmin, which is the most important for this variance derivation. We repeated computations with different $h_\mathrm{ref}$, and they led to quite a similar quality of the fit, but with a slightly better convergence of the residuals for $h_\mathrm{ref}=18$~km. Therefore, in the following we used this parameter value as the best, and the $I_{\mathrm{F},{i}}$ values refer to this distance from hereon.

\begin{figure}[tbh]
\resizebox{\hsize}{!}{\includegraphics* [width=\linewidth]{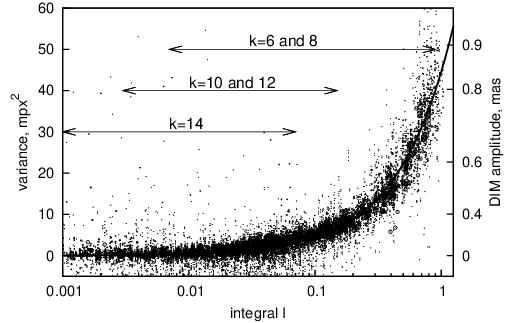}}
\caption{Measured  variance ${\sigma}^2_{\mathrm{a},i}(k,R)$ at zenith for the 30~s exposure as a function of $I_\mathrm{F}$ metric computed at $h_\mathrm{ref}=18$~km for all $k$ modes mixed and a full  data set  (dot sizes show the star brightness). A linear Eq.\,\ref{eq:sa2}  fit  (solid line) visually coincides with the model  dependence (\ref{eq:sigmod}) based on the turbulence profile. Typical ranges of $I_\mathrm{F}$ for each $k$ are shown.}
\label{atm_I}
\end{figure}

Figure \ref{atm_I} demonstrates the good linearity of the measured variances ${\sigma}^2_{\mathrm{a},i}(k,R)$ on $I_\mathrm{F}$ for a full data set using all $k$ modes. We derived $ \hat{A}_0= 0.715$~mas$^2 = 44.95$~mpx$^2 $ with a formal uncertainty of $\pm 0.002$~mas$^2 =\pm   0.13$~mpx$^2$. This is close, but it is a factor 1.009 over the $A_0$ or a factor $\tau=1.019$ over the $A_{\mathrm{c}}(\langle k \rangle)$ (averaged over $k$) estimates expected from the  model Eq.\,\ref{eq:sigmod}.  In the following, we  recomputed the variances ${\sigma}^2_{\mathrm{a},i}(k,R)$, applying corrections for the offset terms $\mathrm{off}_k$  and $\delta^*_k R^{-2}_i$.

\begin{figure}[tbh]
\includegraphics[width=\linewidth]{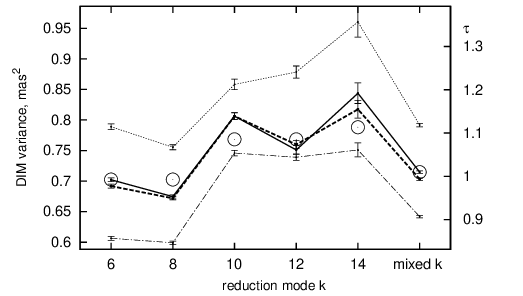}
\caption{DIM variances $  {A}_\mathrm{F}(k) $ (solid) and $\bar{A}_\mathrm{F}(k)$  (dashed lines) derived from FORS2 astrometry at any seeing for each $k$ (and $k$ modes mixed); the  variances  $\tau A_{\mathrm{c}}(k)$  derived from the  model  Eq.\ref{eq:cf} at $\tau=1.019$ are shown by circles. Average of $  {A}_\mathrm{F}(k) $  and $\bar{A}_\mathrm{F}(k)$  for a  good $0.4 - 0.6\arcsec$  and  worse   $0.8 - 1.0\arcsec$ seeing are shown by dash-dotted and  dotted lines, respectively. The right axis $\tau$ is the ratio of DIM variances to $A_0 =0.708$~mas$^2$  estimated with the model Eq.\ref{eq:sigmod} . }
\label{AA_k}
\end{figure}

We proceeded in this investigation as described above but separately for each $k$ reduction mode. We wanted to fit the measured variances ${\sigma}^2_{\mathrm{a},i}(k,R)$ with a linear function,
 \begin{equation}
\label{eq:sa22}
{\sigma}^2_{\mathrm{a},i}(k,R)  =  {A}_\mathrm{F}(k)  I_{\mathrm{F},{i}}(h_\mathrm{ref},k,R),
 \end{equation}
similar to Eq.\,\ref{eq:sa2} using data for each mode $k$ separately. The derived values   (Table\,\ref{coef_A_k}) slightly deviate from  $ \hat{A}_0$, and the deviation pattern correlates well with $A_{\mathrm{c}}(k)$ in Table\,\ref{acf}. We show this in Fig.\,\ref{AA_k}, where  ${A}_\mathrm{F}(k) $ is compared with $\tau A_{\mathrm{c}}(k)$ using a factor $\tau=1.019$ applied to match the model to FORS2 estimates at $k$ averaged. For most $k$,  the inconsistency between both estimates is within three sigma, meaning that the extraction of the DIM variance from FORS2 astrometry is correct, and the conservative accuracy of the coefficients $A_{\mathrm{F}}(k)$ derivation is better than 2\% when considering the worst mismatch between $\tau A_{\mathrm{c}}(k)$ and $ {A}_\mathrm{F}(k) $. { These conclusions also remain valid for von Karman turbulence (Sect.\,\ref{vonM})}.

\begin{table}[tbh]
\caption [] {FORS2 variances  $  {A}_\mathrm{F}(k) $,   $\bar{A}_{\mathrm{F}}(k)$ [mas]$^2$, and a scaling factor $\bar{\tau}(k)$ defined, respectively, by Eq.-s\,\ref{eq:sa2}, \ref{eq:Ak}, \ref{eq:dircmp}. }
{\small
\centering
\begin{tabular}{@{}c@{ }|c@{  \, }c@{ \,   }c@{ \,   }c@{ \,   }c@{ \,   }c@{}}
\hline
\hline
parameter          &\multicolumn{6}{c}{  $k$ / $\beta$}     \rule{0pt}{11pt}\\
\cline{2-7}
              &      6 / 2        &    8 / 3         & 10 / 4      & 12 / 5       &  14 / 6    & mixed modes \rule{0pt}{11pt}\\
\hline             
 $ {A}_\mathrm{F}(k) $     &  0.702    & 0.675    &   0.806     &   0.751  &   0.844   & 0.715 \rule{0pt}{11pt} \\
 $\bar{A}_{\mathrm{F}}(k)$    &  0.692 &   0.672 &   0.806 &   0.760  &  0.817  & 0.703   \\
    $\bar{\tau}(k)$  & 1.004  & 0.975 & 1.069 & 1.008 & 1.057 & 1.002 \\
\hline
uncert.-y  &  $\pm$0.003     &$\pm$0.003    &  $\pm$0.006     &  $\pm$0.008   &   $\pm$0.013 & 0.002  \rule{0pt}{11pt} \\
\hline
\end{tabular}           
\label{coef_A_k}
}
\tablefoot{The uncertainties are valid for each term.}
\end{table}

\subsection{Direct comparison of the observed and atmospheric turbulence model  variances}{\label{on_int2}}
We exploited another yet slightly different possibility in order to directly compare the measured ${\sigma}^2_{\mathrm{a},i}(k,R)$   variances for each P-star to those expected from the turbulence model. For that purpose,  we computed  the DIM variance at zenith for the adopted atmospheric model for each P-star, which in accordance with  Eq.\ref{eq:cf} is $\sigma^2_{\mathrm{F},i}(k,R)  =        \mu_{i,k,R}  \sigma_\mathrm{c}^2(k,R) $  at $T=T_0$, and using  terms $\mu_{i,k,R}$ individually  for each star. Then we assumed that the observed ${\sigma}^2_{\mathrm{a},i}(k,R)$ and the model $\sigma^2_{\mathrm{F},i}(k,R)$ variances are equal with an accuracy to the scaling fit parameter $\bar{\tau}(k)$: 
 \begin{equation}
\label{eq:dircmp}
{\sigma}^2_{\mathrm{a},i}(k,R)= \bar{\tau}(k) \sigma^2_{\mathrm{F},i} (k,R).
 \end{equation}
Thus, the corrected estimate of the DIM amplitude in Eq.\ref{eq:ack}, which now matches the FORS2 observations for each $k$ separately, is  
 \begin{equation}
\label{eq:Ak}
\bar{A}_{\mathrm{F}}(k)=\bar{\tau}(k) A_{\mathrm{c}}(k).
 \end{equation}
The fit parameters  $\bar{\tau}(k)$ and the corresponding DIM amplitudes  $ \bar{A}_{\mathrm{F}}(k) $ computed with Eq.\ref{eq:Ak} are given in Table\,\ref{coef_A_k}. The amplitudes match well with the $ {A}_\mathrm{F}(k)$ derived also from FORS2 data  (Fig.\,\ref{AA_k}) but  differ over three sigma from the ${\tau}(k) A_{\mathrm{c}}(k)$ correspondent to the turbulence model. We consider that the possible reason of this discrepancy is a minor variability of $\mu_{i,k,R}$  with $d$, which we neglected in the computation of  both values, and of course the inconsistency between the actual and the model turbulence. In any case, the relative discrepancy is below about 3\% and is thus sufficient for practical purposes.

\begin{figure}[tbh]
\includegraphics[width=\linewidth]{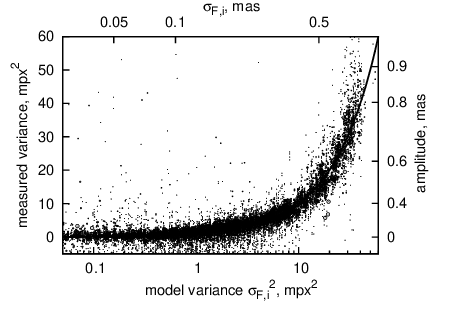}
\caption{Same as in Fig.\,\ref{atm_I}  but the horizontal axis refers to the DIM variance (\ref{eq:cf}) based on the vertical turbulence profile and using $I$  for each P-star. The fit function (\ref{eq:dircmp}) with parameter $\bar{\tau}(\langle k \rangle)$ is shown with a solid line.}
\label{HHAA}
\end{figure}
   
Figure \ref{HHAA}  compares the observed ${\sigma}^2_{\mathrm{a},i}(k,R)$ with the model $\sigma^2_{\mathrm{F},i}(k,R)$ variances for all $k$ modes. It also shows a linear dependence (\ref{eq:dircmp}) with the parameter $\bar{\tau}(\langle k \rangle)=1.002 \pm 0.002$ equal to the average of $\bar{\tau}(k)$ over all $k$. This parameter value is comparable at five sigma to the  ratio  $\tau=1.019 \pm 0.002$ between FORS2 and the model data obtained as described above in this Section. Furthermore, with Eq.\ref{eq:Ak}, we derived $\bar{A}_{\mathrm{F}}(\langle k \rangle) =0.703\pm 0.002$~mas$^2$ (= 44.2~mpx$^2$) near the parameter $A_0$ value defined by  Eq.\ref{eq:turb}.
  
Considering that the coefficients $  A_\mathrm{F}(k) $ and  $\bar{A}_{\mathrm{F}}(k)$ in Table\, \ref{coef_A_k} are little dependent on $k$, we found a quite accurate approximation,
\begin{equation}
\label{eq:slin}
\sigma_{\mathrm {a}} \approx (860 \pm 40) \mu{\mathrm {as}} \sqrt{I(k,R)} (T_0 /T)^{0.5},
\end{equation}
valid for polynomial orders $\beta$ from two to six. Here, $T_0=30$~s and the metric integral $I(k,R)$ is computed at $h_\mathrm{ref}=18$~km. This equation allows one to roughly estimate $\sigma_{\mathrm {a}}$ without any computations of the integral $I$, instead taking its approximate value from Fig.\,\ref{tr}. 

As an additional control, we inspected the sample of the least-squares fit of  Eq.\,\ref{eq:sa2} residuals. If the data treatment is correct,  they should have a Gaussian distribution with a sigma parameter: 
 \begin{equation}
\label{eq:srez}
\varepsilon=E \hat{\sigma}_\mathrm{mod}^2  [F(h_\mathrm{ref})/F(h_\mathrm{ref}\sec{z})](\sec{z})^{-\nu-p-1}(T/T_0),
\end{equation}
where the $E$ values are those already derived in Sect.\,\ref{internal} for each field and are of the order of a unit. We find that after normalization with $\varepsilon$, they are distributed approximately by a normal law, with a moderate asymmetry in the histogram of the residuals. Further, the rms of the normalized residuals is equal to 1.03, near the expected unit value.

We also tested the stability of the $ {A}_\mathrm{F}$ and  $\bar{A}_{\mathrm{F}}$ terms with seeing by splitting data into a few narrow FWHM ranges, and we found a moderate increase of the DIM variance  by about $20\%$ when comparing the best seeing (FWHM= 0.4 -- 0.6\arcsec) to the worst (0.8 -- 1.0\arcsec) seeing (Fig.\ref{AA_k}). Thus, they are comparable in spite of a twofold seeing variation. Hypothetically, this increase can be caused by a weak correlation between the FWHM and the $C_n^2$ at high altitudes, which  leads to the observed effect. Some evidence of an actual correlation between these quantities has been found by \citet{SCIDAR1}, who formed average  $C_n^2$ profiles over Paranal for different classes of atmospheric conditions, particularly when the turbulence concentrated above the ground layer is high or when it is low along any altitudes. The value of   $C_n^2$ in the first case is a factor of three over that in the second case \citep[][Fig.9]{SCIDAR1} for altitudes 5 -- 13~km. This is also reflected in a nearly twofold-different median seeing,   correspondingly   0.76\arcsec and 0.46\arcsec, which is quite near the FWHM ranges we tested. We numerically verified that the 5 -- 13~km altitudes  produce about 8\% of the DIM variance; thus, the twofold increase of  $C_n^2$ at these altitudes is likely sufficient to explain the 20\% change in the DIM variance considered. However, we also admit the possibility that the detected change of the DIM variance with seeing is false and caused by an insufficiently good reduction model. In this case, we should restrain ourselves to the results derived for the average seeing.

In this section we have demonstrated that the measured DIM variance matches that expected for the turbulence profile over a wide (three decades) range of the metric integrals' change and data represented by  the mix of parameters $R$, $k$,  for observations obtained under various star distributions in sky fields and variable seeing. We conclude the following: 
\begin{itemize}
\item  The astrometric reduction of the FORS2 observations, the calibration of error terms, and the extraction of the DIM variance were performed sufficiently well;
\item  The FORS2 data, treated with different approaches, produced estimates of the DIM amplitude in zenith in the range of 0.705 -- 0.722~mas$^2$ for a 30~s exposure and a reference field with a unit integral $I$. This is 0.995 -- 1.019 of the value based on the turbulent vertical profile if it was preliminary scaled by a factor $\tau_0=0.84$;
\item The DIM variance is stable to within $\pm 20\%$ of its variations with respect to the seeing change.
\end{itemize}

\subsection{Dependence of the DIM variance on the  field size}{\label{s_tr}}

The significant variability of the metric integral $I_{\mathrm{F},i}$ between the individual P-stars and star fields is a problem when we consider the dependence of the measured ${\sigma}^2_{\mathrm{a},i}(k,R)$ on $R$ or  $T$. This is because ${\sigma}^2_{\mathrm{a},i}\sim I $ above the effect propagates to a significant scatter of individual DIM measurements at a fixed $R$ or $T$, and it can be erroneously interpreted as incorrect data processing. Moreover, it can bias the model parameter estimates. Of course, this is of no importance for the modeled dependence of  ${\sigma}^2_{\mathrm{a},i}(k,R) $ on  $I$, whose variations are directly ingested by a linear function Eq.\,\ref{eq:sa2}. In the following, we tried to reduce these  problems. 

\begin{figure}[tbh]
\includegraphics[width=\linewidth]{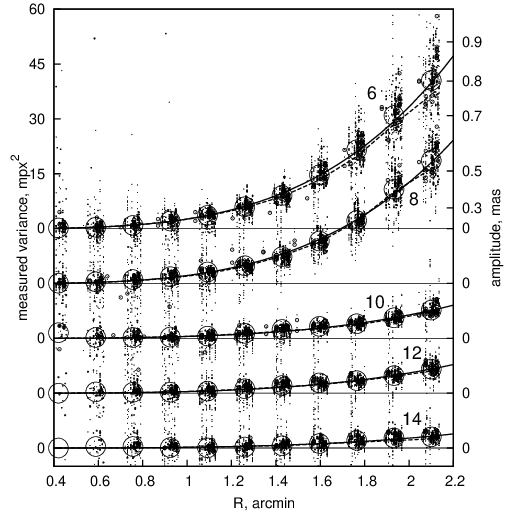}
\caption{Each star variance  ${\sigma}^2_{R,i}(k,R)$ as a function of field size $R$  (dots; sizes are proportional to star brightness) for each reduction mode, their average $\langle {\sigma}^2_{R,i}(k,R) \rangle $ for each $R$  (large circles), fit functions Eq.\,\ref{eq:hh_r} (solid), and the DIM variance $\sigma^2_{\mathrm{F}}(k,R) $ (Eq.\,\ref{eq:Ac}) based on the atmospheric model (dashed lines). For better visualization, the plots for each $k>6$ are shown with a negative offset. The dots are dispersed horizontally for different star fields. }
\label{fg_R}
\end{figure}

\begin{figure}[tbh]
\includegraphics[width=\linewidth]{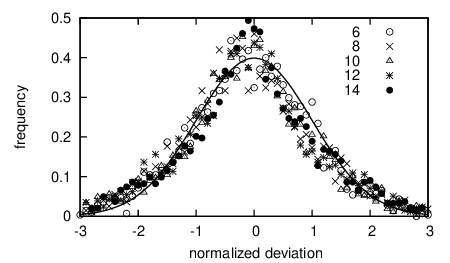}
\caption{Frequency distribution of the normalized deviations ${\sigma}^2_{R,i}(k,R) - \langle {\sigma}^2_{R,i}(k,R) \rangle $ for a full sample of DIM probes at each reduction mode $k$ (different signs). The normal distribution is shown with a solid line. }
\label{fg_demo}
\end{figure}

As commented, because the ${\sigma}^2_{\mathrm{a},i}$ variances  (\ref{eq:snorm})  are not exactly comparable in the domain of $R$, we converted them to normalized quantities (Fig.\,\ref{fg_R}),
 \begin{equation}
\label{eq:s_norm}
{\sigma}^2_{R,i}(k,R)=\langle \mu_{k,R} \rangle {\sigma}^2_{\mathrm{a},i}(k,R) I_\mathrm{c}(h_\mathrm{ref},k,R)/ I_{\mathrm{F},i}(h_\mathrm{ref},k,R), 
 \end{equation}
that are free from variations of values  $I_{\mathrm{F},i}(h_\mathrm{ref})$ for individual stars and rather represent the expectation of the DIM variance  $\sigma^2_\mathrm{a}$ for a typical star field in the Galactic plane at a given $R$ and $k$. Because in this study the data was split into 11 discrete sizes of $R$,  within each $R$, we averaged the individual $i$-star estimates Eq.\,\ref{eq:s_norm} to   $\langle {\sigma}^2_{R,i}(k,R) \rangle $, thus deriving well-defined estimates of the DIM variances  $\sigma^2_\mathrm{a}$  shown in Fig.\,\ref{fg_R} by circles. At this step, we expected that individual estimates of  ${\sigma}^2_{R,i}(k,R)$ are distributed as Gaussian   values with a mathematical expectation $\langle {\sigma}^2_{R,i}(k,R) \rangle $ and a sigma parameter ${\varepsilon}_0(k,R) =\varepsilon \langle \mu_{k,R} \rangle  I_\mathrm{c}(h_\mathrm{ref},k,R)/ I_{\mathrm{F},i}(h_\mathrm{ref},k,R)$, with $\varepsilon$  defined by Eq.\,\ref{eq:srez}.  We formed a histogram   of normalized deviations  ${\sigma}^2_{R,i}(k,R) -  \langle {\sigma}^2_{R,i}(k,R) \rangle $ for the data accumulated over $R$  (each $k$ mode was handled separately) and compared it with the normal law (Fig.\,\ref{fg_demo}). The observed distribution is slightly asymmetric, but it has a width parameter of approximately one unit (0.99 on average). We consider that this match with a normal distribution is sufficiently good, which suggests the level of correctness of the data processing.

To model the dependence of the observed variances ${\sigma}^2_{R,i}(k,R)$  on $R$,  we fit them, for each $k$ separately, by the expression
 \begin{equation}
\label{eq:hh_r}
\hat{\sigma}^2_{R,i}(k,R) = \hat{a}_{{R}}^2(k) F^2_{\mathrm{eff}}(k,R) (R/R_0)^{\nu +p}
\end{equation}
with a single free parameter $\hat{a}_{{R}}(k)$. By its structure, the function (\ref{eq:hh_r}) is equivalent to  Eq.\,\ref{eq:Ac}, both representing the model fitting of the measured ${\sigma}^2_{R,i}(k,R)$ or the turbulence-based variances  $\sigma^2_{\mathrm{F}}(k,R) $ via expressions valid for narrow fields.  The terms $\hat{a}_{{R}}^2(k)$ are therefore the measured equivalent of the model amplitudes $a_{\mathrm{F}}^2(k)$ in (\ref{eq:Ac}).  For computations, we assumed that $F^2_{\mathrm{eff}}(k,R)=1$ is a good approximation, which requires only some  small second-order corrections applied later.

The amplitudes $\hat{a}_{{R}}(k)$ derived  for each reduction $k$ mode (with $\chi^2 =1.07$  reduced residuals of the fit) are  given in Table \ref{dep_r} for exposure $T=30$~s and $R_0=1\arcmin$. Computations were made for the data with any seeing and separately for good and bad seeing. We found an about 10\% increase of the amplitude in the observations under bad seeing conditions, which agrees with a similar 20\% increase of the DIM variances found in Sect.\ref{on_int2}. 

\begin{table}[tbh]
\caption [] {Coefficients $\hat{a}_{{R}}(k)$ in Eq.\,\ref{eq:hh_r}  and $\hat{a}_{{T}}(k)$ in Eq.\,\ref{eq:s00}  derived from FORS2 observations,  with nominal  uncertainties.}
{\centering
\begin{tabular}{@{}c@{\,}|c@{ \,}c@{ \,}c@{ \,}c@{ \,}c@{}}
\hline
\hline
FWHM     &\multicolumn{5}{c}{  $k$ / $\beta$}    \rule{0pt}{11pt}\\
\cline{2-6}
arcsec         & 6 / 2   & 8 /3      &10 / 4    & 12 / 5   & 14 / 6   \rule{0pt}{11pt} \\
\cline{2-6}
         &\multicolumn{5}{c}{ $\hat{a}_{{R}}(k)$ [$\mu$as] } \rule{0pt}{11pt} \\
\hline         
0.4--0.6     & 192.6$\pm0.3$   &  174.4$\pm0.3$  & 85.6$\pm0.2$   & 80.3$\pm0.2$   & 55.6$\pm0.4$  \rule{0pt}{11pt}\\
0.8--1       & 215.6$\pm$0.5    & 199.7$\pm$0.4  &94.3$\pm$0.4     &88.4$\pm$0.4   &63.3$\pm$0.6    \\
any         & 204.2$\pm$0.4    & 186.0$\pm$0.3  &89.4$\pm$0.2     &83.2$\pm$0.3   &58.3$\pm$0.4    \\
\hline                                     
         &\multicolumn{5}{c}{ $\hat{a}_{{T}}(k)$ [$\mu$as]} \rule{0pt}{11pt} \\
\hline         
0.4--0.6    & 191.1$\pm0.3$   &  176.5$\pm0.3$  & 83.3$\pm0.1$   & 76.6$\pm0.1$   & 52.6$\pm0.2$ \rule{0pt}{11pt} \\
0.8--1      & 218.1$\pm$0.5    & 202.9$\pm$0.4  &93.6$\pm$0.2     &86.8$\pm$0.2   &60.3$\pm$0.3    \\                   
any            & 205.0$\pm$0.4    & 188.2$\pm$0.3  &88.7$\pm$0.2     &82.4$\pm$0.2   &55.8$\pm$0.2    \\
\hline                                     
 &\multicolumn{5}{c}{ $a_{\mathrm{F}}(k)$ [$\mu$as]} \rule{0pt}{11pt} \\
 \hline
any   & 204$\pm 4$  &   185$\pm 4$ & 90$\pm 2$ &  82$\pm 1$ &  55$\pm 1$  \rule{0pt}{11pt}\\
  \hline                                     
\end{tabular}           
\label{dep_r}
}
\tablefoot{Coefficients were derived for good, bad,  any seeing conditions,   exposure $T=30$~s and $R_0=1\arcmin$.}
\end{table}

In addition, we  modified the model Eq.\,\ref{eq:hh_r}, introducing an extra free fit parameter instead of the constant order $\nu +p=3.67$. We derived the estimates 3.73 $\pm$ 0.06,  3.75 $\pm$ 0.06, 3.39 $\pm$ 0.07,  3.74 $\pm$ 0.07, and  3.91 $\pm$ 0.31, listed in ascending order of $k$ from six to 14, correspondingly. The average is 3.65, which actually coincides with the 3.67 expected for the Kolmogorov turbulence. The only over three-sigma deviation is seen  for $k=10$,  probably due to a high (0.97) correlation between the model parameters $\hat{a}_{R}$ and $\nu +p$. Thus, our data support the Kolmogorov type of the turbulence at altitudes 16--18~km.

Using the coefficients  $\hat{a}_{R}^2(k)$ in Table\,\ref{dep_r} and Eq.\,\ref{eq:hh_r},  it is easy to find the DIM variance $\sigma_{\mathrm a}$ for any field size $R$. However, we should note a problem, namely that the coefficients  $\hat{a}_{R}^2(k)$  in Eq.\,\ref{eq:hh_r}  were obtained with approximation of $F^2_{\mathrm{eff}}(k,R)=1$ because these terms are unknown. In order to handle this problem, we exploited CSD as a good imitator of the FORS2 fields properties easily modeled for any $R$. This observation allowed us to note a similar problem for CDS fields when the variances $\sigma^2_{\mathrm{c}}(k,R)$ in Eq.\,\ref{eq:calc} were fit by the model Eq.\,\ref{eq:Ac} under the approximation $F^2_{\mathrm{eff}}(k,R)=1$. In this case, the quantity $\hat{ F}^2(k,R) = \sigma^2_{\mathrm{c}}(k,R) / \hat{\sigma}^2_{\mathrm{c}}(k,R)$ applied as a correcting factor to $\hat{\sigma}^2_{\mathrm{c}}(k,R)$ in a trivial way removes the mismatch between the functions referred to here for the CSD fields, allowing one to restore $\sigma^2_{\mathrm{c}}(k,R)$ values via parameters $a_{\mathrm{c}}^2(k)$. Quite similarly, calibration of $\hat{\sigma}^2_{R}(k,R)$ with this factor removes the problem for FORS2 fields. Of course, the terms $\hat{ F}^2(k,R)$ we estimated at $R_0$ are exactly equal in both fitting expressions (\ref{eq:calc}) and (\ref{eq:hh_r}), that is at  $R_0 =1\arcmin$.  The computed values $\hat{ F}(k,R)$ are all  $\pm 0.05$ near to a unit, except for the smallest value $R=0.5\arcmin$, where some deviations mount to $\pm 0.13$. It follows that use of Eq.\,\ref{eq:hh_r}, which does not contain a factor $F^2_{\mathrm{eff}}$, is quite sufficient to derive the DIM variance with an accuracy to $\sim 5$\%.

	To find more accurate estimates of DIM amplitudes, we used  $\hat{F}(k,R)$ as a {scaling calibrator to $(\hat{a}_{{R}}(k)+\hat{a}_{{T}}(k))/2$ } and computed  $\sigma_{\mathrm a}$  for a few field sizes (Tabl.\,\ref{at_1min}). Due to the large quantity of statistics used, the {formal } uncertainty of these estimates is about 1\% only. However, we note that for a single randomly selected sky field, even in the Galactic plane, we expect to encounter strong variations of $\mu_{i,k}$ due to the peculiar distribution of the stars, leading to variations of the metric integral and DIM variance. {Therefore, in Table\,\ref{at_1min} we give uncertainties that treat typical fluctuations of star distribution in  fields expressed by  $\mu_{i,k}$  as a random event. } This is useful to predict ${a}_{{R}}(k,R)$ by guessing for a randomly selected field near the Galactic plane without taking into consideration the details of star distribution. Numerically, this type of uncertainty exceeds the formal uncertainty very much, roughly by a factor of ten to 20. In many cases, it mounts to $\sim$20\% of the $\sigma_{\mathrm a}$ value, or even more. 

\begin{table}[tbh]
\caption [] {Differential image motion amplitude  $\sigma_{\mathrm a}$ [$\mu$as] at 30~s exposure in the zenith for a typical FORS2 star field in the Galactic plane. }  
{\centering
\begin{tabular}{@{}c|rrrrr@{}}
\hline
\hline
$R$     &\multicolumn{5}{c}{   $k$ / $\beta$  }    \rule{0pt}{11pt}\\
\cline{2-6}
          & 6 / 2   & 8 /3      &10 / 4    & 12 / 5   & 14 / 6   \rule{0pt}{11pt} \\
\hline                                     
0.5\arcmin &   58$\pm$26 &  47$\pm$10 &  28$\pm$13 &  25$\pm$10 &  17$\pm$ 8 \rule{0pt}{11pt}\\
1\arcmin   &  206$\pm$42 & 179$\pm$27 &  90$\pm$31 &  85$\pm$18 &  59$\pm$22 \\
1.5\arcmin &  423$\pm$37 & 384$\pm$26 & 194$\pm$29 & 181$\pm$26 & 129$\pm$37  \\
2.0\arcmin &  734$\pm$44 & 670$\pm$36 & 318$\pm$33 & 295$\pm$28 & 200$\pm$36  \\
\hline
\end{tabular}           
\label{at_1min}
}
\tablefoot{  Uncertainties indicate the scatter of the individual estimates for  P-stars in FORS2 fields.}
\end{table}

Considering a decrease of $\hat{a}_{{R}}(k)$ for high $k$ and low $R$, one can expect significant improvement of the astrometric accuracy when applying the extreme values of these parameters. However, its value, modeled by the quantity  $\sigma_\mathrm{mod}$ in  Eq.\,\ref{eq:sig_oc0}, is the sum of all noise components. Therefore, it is a bivariate function of $k$ and $R$. The interplay between these variables is not obvious and depends on the peculiarity of the observations. In particular, while $\sigma_\mathrm{r}$ decreases with the increase of $R$, the component $\sigma_\mathrm{a}$ follows a reverse dependence. Hence, the minimum of $\sigma_\mathrm{mod}$ occurs at some optimal $R$ size where  $\sigma_\mathrm{r} =  \sigma_\mathrm{a}$  \citep{Lazorenko2006}. For our observations, this  value is usually between 0.5 and 1.5\arcmin, depending on $k$ and the actual star distribution. A curious feature is that use of different sets of $k$ and $R$ often lead to a quite moderate ($\sim$20\%)  improvement in $\sigma_\mathrm{mod}$.

\subsection{Dependence of the DIM variance on  exposure}{\label{exp}}

\begin{figure}[tbh]
\resizebox{\hsize}{!}{\includegraphics* [width=\linewidth]{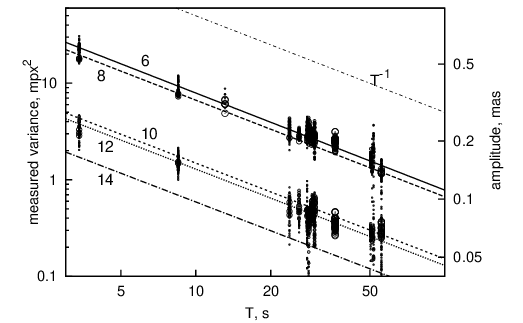}}
\caption{Dependence of the DIM variance  $\sigma^2_{T,i}$ reduced to $R=1\arcmin$ field size on the exposure $T$. The figure shows the individual estimates for each star (circle sizes are proportional to the star brightness) for the reduction with $k=6$  (upper symbol sequence) and $k=12$ (lower sequence); fit functions $T^{B(k)}$ (lines) for each $k$; theoretical dependence $T^{-1}$. Dots  are clustered in groups according to  exposure in  sky fields.}
\label{R_T}
\end{figure}

We tested also a well-known inverse dependence, Eq.\,\ref{eq:sa}, of  $\sigma^2_{\mathrm{a}} $ on  $T$  for long  exposures. For that purpose, we reduced the variances $ \left. {\sigma}^2_{\mathrm{a},i}\right|_{z=0}$ computed with  Eq.\,\ref{eq:snorm0} and related to zenith to the variances
\begin{equation}
\label{eq:s0}
{\sigma}^2_{T,i}(k) = \left.\langle \mu_{k,R} \rangle     {\sigma}^2_{\mathrm{a},i}\right|_{z=0}    I_\mathrm{c}(h_\mathrm{ref},R)/ I_{\mathrm{F},i}(h_\mathrm{ref},R)  (R_0/R)^{\nu +p}.
\end{equation}
In this way, due to a factor $ (R_0/R)^ {\nu +p}$, the data  obtained with several field radii $R$ were converted to some common field size $R_0$. For verification, we used all the measurements available at $R>1.6$\arcmin, where the DIM  signal is highest, thus converting most of the accumulated data to conditions that differ by exposure $T$ only. Values $\sigma^2_{T,i}$ obtained with $R_0=1\arcmin$ and that refer to a typical star field at the Galactic equator were fit by the  model
\begin{equation}
\label{eq:s00}
{\sigma}^2_{T,i}(k) =\hat{a}_{T}(k)^2 (T_0/T)^{B(k)}
\end{equation}
with two free parameters $\hat{a}_{T}(k)$ and $B(k)$  for each $k$ mode with $T_0=30$~s  (typical exposure in our observations). By its structure, Eq.\,\ref{eq:s00} and the model expression Eq.\,\ref{eq:Ac} are identical if $F^2_{\mathrm{eff}}(k,R)=1$. For each $k$, the observed power indexes $B(k)$  were found to be near the expectation, $B=-1$  (Fig.\,\ref{R_T}), varying between $-0.990 \pm 0.007$ (for $k=14$) and $-1.005 \pm 0.004$ (for $k=6$) and thus  statistically equal to  $-1$. The result $B=-1$ means that there is a small only  input of components with a variance that does not depend on $T$; otherwise. we would register the deviation of $B(k)$ from $-1$. In particular, this refers to geometric distortions of high orders  (Sect.-s\,\ref{filt},\,\ref{obs}), at least those that vary between exposures. Consequently, our estimates of the measured DIM variance  ${\sigma}^2_{\mathrm{a},i}$ and all the derived parameters, for example $\hat{a}_{R}(k)$, are not biased by these effects.

We proceeded by adopting $B=-1$ in Eq.\,\ref{eq:s00} and recomputed the amplitude coefficients $\hat{a}_{T}(k)$ for each $k$ mode. The fit values given in Table\,\ref{dep_r} are nearly the same for the mode pairs (6, 8) and (10, 12), as expected according to the discussion in Sect.\,\ref{stand}; this comment refers to  $\hat{a}_{R}(k)$ also. A comparison between  $\hat{a}_{R}(k)$ and $\hat{a}_{T}(k)$, both related to  $R_0=1\arcmin$ and $T=30$~s, revealed a divergence between some estimates over the three-sigma nominal uncertainty. This means that the uncertainties are probably underestimated by a factor of 3.8. The conservative and more real error estimate is therefore roughly $\pm 1$~$\mu$as, or yet comfortable enough $\sim \pm 1$\% for any value in Table\,\ref{dep_r}.

For clarity, in Table\,\ref{dep_r} we reproduce the estimates of $a_{\mathrm{F}}(k)$ from Table\,\ref{acf} based on the turbulence profile, our methodology of DIM estimation using the spectral approach, and the $\langle \mu_{k,R} \rangle $ terms  (the ratio between integrals $I$ for FORS2 and for CSD). In spite that $a_{\mathrm{F}}(k)$ terms were  obtained quite differently and irrespective of the actual observations, a good match of the model predicted with measured variances $\hat{a}_{R}(k)$ and $\hat{a}_{T}(k)$ is evident.

\subsection{Change of DIM variance across the field of view}{\label{across}}

We especially emphasize that both the model and the observed DIM variances discussed in this article refer to a single target object indexed $i=0$ near the field center, which corresponds to astrometry focused on precise measurement of a single binary system object. The estimate of DIM variance in other field points is of interest in works focused, for example, on astrometry of stellar clusters when science objects are spread over the field. We found the solution to this problem for a star at distance $r$ off field from the center using the spectral description of the DIM process.

In the case concerned, we are interested in estimating the DIM for off-center stars indexed as $i_{\mathrm{off}}>0$  under the condition of using the least-squares transformation to the reference frame identical to that used for the reduction of the central $i=0$ object, that is using the same projective matrix $\boldsymbol{a}$ (Sect.\,\ref{filt}). The filter $Q(q)$ for the relative shift of $i_{\mathrm{off}}$ star is computed with coefficients $a_{i}$ in Eq.\ref{eq:L2} equal, in this case, to  elements $a_{i_{\mathrm{off}},i}$ in the line $i_{\mathrm{off}}$ of the matrix $\boldsymbol{a}$. Also,  $\rho_i $ in Eq.\ref{eq:L2} are now the angular distances between $i_{\mathrm{off}}$ and the other $i$ star. To derive the numerical solution, we adopted the CSD model, which  matches well with the star distributions at the Galactic plane and the computed metric integrals $I_r(k)$ as a function of off-center separation $r$. We found that $ I_r(k)$ for $r$ till $0.1R$ is within $\pm 20 \%$ of the $I_{r=0}(k)$ value in the field center. For off-center separations 0.3--0.5 $R$ and $k=6$, 10, 14, the excess of  $ I_r(k)$ over $I_{r=0}(k)$ increases respectively to 2.5, 1.5, and 1.2. Consequently, the DIM variance follows this change.

\section{DIM in astrometric observations with other large seeing-limited telescopes}{\label{cn2}}

Using the methodology of DIM estimation based on the spectral description of this process and the availability of the vertical turbulence structure, we can compute variances $\sigma^2_{\mathrm{a}}$ expected for any other telescope location with any aperture $D$. We did it for the telescope of VLT size at the sea level and for the ELT at its actual location. In the first case, using  Eq.\,\ref{eq:turb} and adopting the same turbulent profile as for VLT, we obtained $A_0= 79.1$~mpx$^2= 1.26$~mas$^2$, or a factor 1.78 over that  for VLT. Therefore, at the sea level for the VLT-size telescope, values ${a}_R(k,R)$ are expected to be 1.3 times higher than those in Tabl.\,\ref{at_1min}.

Similar computations for the ELT located at 3046~m above the sea level and only 30~km away from VLT produced $A_0= 41.2$~mpx$^2 =0.655$~mas$^2$, or a factor 0.92 better than for VLT. Of course, the main improvement in the DIM variance follows from its large aperture $D=39$~m due to the dependence $\sigma^2_{\mathrm{a}} \sim D^{-\nu}$, in accordance with Eq.\,\ref{eq:L4}.  Using this dependence and adopting new $A_0$ value, we scaled  ${a}_R(k,R)$ from Tabl.\,\ref{at_1min} known for VLT and thus derived the DIM variance for ELT (Tabl.\,\ref{ELT}). These amplitudes are on the order of ten better in comparison to those for VLT. In particular, we find that roughly ${a}_R \approx 2 - 5$~$\mu$as for the $R=0.5\arcmin$ field if transformation to the reference frame is made with  orders $k$ of the reduction polynomials from six to 14. 

\begin{table}[tbh]
\caption [] {Same as in Tabl.\,\ref{at_1min} for ELT,  at Galactic equator, with {the expected scatter of amplitudes for a single randomly selected field}.}
{\centering
\begin{tabular}{@{}c@{\,}|@{\,}rrrrr@{}}
\hline
\hline
$R$     &\multicolumn{5}{c}{   $k$ / $\beta$  }    \rule{0pt}{11pt}\\
\cline{2-6}
          & 6 / 2   & 8 /3      &10 / 4    & 12 / 5   & 14 / 6   \rule{0pt}{11pt} \\
\hline         
0.5\arcmin &   5.4$\pm 2.4$ &  4.4$\pm$1.0 &  2.6$\pm$1.2 &  2.4$\pm$0.9 &  1.6$\pm$0.8\rule{0pt}{11pt}\\
1\arcmin   &  19.1$\pm$3.9 & 16.6$\pm$2.5 &  8.3$\pm$2.9 &  7.9$\pm$1.7 &  5.5$\pm$2.1\\                
1.5\arcmin &  39.3$\pm$3.4 & 35.6$\pm$2.4 & 18.0$\pm$2.7 & 16.8$\pm$2.4 & 12.0$\pm$3.4\\                
2.0\arcmin &  68.1$\pm$4.1 & 62.2$\pm$3.3 & 29.5$\pm$3.0 & 27.4$\pm$2.6 & 18.5$\pm$3.3\\                
\hline
\end{tabular}           
\label{ELT}
}
\end{table}

\section{Conclusion}{\label{conc}}

Astrometric reduction uses the least-squares transformation to the reference frame via coordinate polynomials of the order $\beta$, which is necessary to reduce geometric distortion of the field. At the same time, it mitigates the DIM effect caused by the atmospheric turbulence. We developed a model that adequately describes the image motion in the spectral domain, where instead of $\beta$, the use of the mode parameter $k$ (Eq.\,\ref{eq:beta}) is more convenient. This model allows us to  translate the turbulence profile, if known, to the amplitude of the DIM displacements $\sigma_{\mathrm{a}}$. A key quantity in this model is the metric integral $I(d)$ (Eq.\,\ref{eq:int}), which numerically represents observational parameters (telescope diameter, field size, exposure, zenith angle, star distribution, brightness, seeing, background) as well as reduction parameters ($\beta$ or $k$) and is computed for the distance $d$ to the turbulent layer.

By prime examination, we have found that the measured (derived from observations) and the turbulent model DIM variances are best comparable when the latter are scaled roughly by a factor of $\tau_0=0.84$, which is not a unit, probably due to the uncertainty in the turbulence profile. With this correction, the numeric value of $\sigma_{\mathrm{a}}^2$ in zenith at a 30~s exposure is equal to the parameters $A_{\mathrm{c}}(k)$ in Table\,\ref{acf} for an imaginary star field with a unit $I(d)$ computed at an effective distance $d=18$~km to the turbulent layers. Values of $A_{\mathrm{c}}(k)$  depending on $k$ (or $\beta$)  vary between 0.69 and 0.77~mas$^2$. 

We ran several tests to compare the model predictions with the observations. In the first test, we fit FORS2 measured DIM variances by a linear function of $I$ and derived coefficients $A_{\mathrm{F}}(k)$, which by definition should be equal to $A_{\mathrm{c}}(k)$, but which are the observed quantities not dependent of the turbulence profile. The coefficients derived with FORS2 vary in the comparable range of 0.67--0.84~mas$^2$ (Table\,\ref{coef_A_k}), depending on $k$ also, with a clear correlation between both results (Fig.\,\ref{AA_k}, Sect.\,\ref{on_int1}). Thus, the model $A_{\mathrm{c}}(k)$ and the observational $A_{\mathrm{F}}(k)$ estimates are quite close.  

In the second test, the measured variances were compared directly with those expected from the turbulent model (Fig.\,\ref{HHAA}) and fit by a linear dependence with a scaling coefficient $\bar{\tau}(k)$. We derived that $\bar{\tau}(k)$ is in the range of 0.995--1.019 (Sect.\,\ref{on_int2}). Thus, the correlation between the two estimates is excellent. A good agreement of the observations with a model is supported by the fact that the residuals of the least-squares fit obtained for these two tests have a Gaussian distribution with an rms very near its model expectation.

For another test, we investigated the dependence of the measured variances on the field size $R$ and found that it follows an expected power dependence $R^{\nu +p}$, which supports the Kolmogorov type of the turbulence (specified by $p=2/3$) at effective altitudes 16--18~km. With this power law, we derived the DIM amplitude $\hat{a}_{{R}}(k)$ for FORS2 at $R=1\arcmin$ field, the zenith, and the 30~s exposure for each $k$ or $\beta$ (Table\,\ref{dep_r}). Using the turbulence profile (not using the actual FORS2 observations), we derived an alternative expression (Eq.\,\ref{eq:Ac}) of a similar structure but with coefficients  $a_{\mathrm{F}}(k)$ estimated using turbulent data. Both coefficients, $\hat{a}_{{R}}(k)$ and $a_{\mathrm{F}}(k)$ shown in Table\,\ref{dep_r}, are comparable to within the uncertainty limits. 

In addition, we analyzed the dependence of the measured DIM variance ${\sigma}^2_{\mathrm{a}}$ on $T$  (Fig.\,\ref{R_T}) and found that  for any $k$, the observed power law is $T^{-1}$, as predicted by the model Eq.\,\ref{eq:s00}. In this test, estimates $\hat{a}_{T}(k)$ of the measured DIM variances at $R=1\arcmin$ and $T=30$~s  match well with both turbulence model coefficients, $a_{\mathrm{F}}(k)$ and $\hat{a}_{R}(k)$, derived from observations in a different way  (Table\,\ref{dep_r}). The observational fact that ${\sigma}^2_{\mathrm{a}}$ follows an expected power law  $\sim T^{-1}$ means, furthermore, that the input of the static (time-constant) field deformations is negligible in comparison to that of the DIM for any $k$ or $\beta$ applied.

All of these tests demonstrated the agreement, {to $\sim 1\%$ }, between the observational and the model lines of the investigation. We emphasize that the spectral-based model is not limited to FORS2 only and that it can be generalized to predict $\sigma^2_a$ for other very large telescopes, including ELT (Sect.\,\ref{cn2}). Of course, our model estimates are restricted to telescopes that do not use adaptive optics. 

\bibliographystyle{aa}
\bibliography{atm_bib} 

\begin{thebibliography}{28}
\expandafter\ifx\csname natexlab\endcsname\relax\def\natexlab#1{#1}\fi

\bibitem[{{Appenzeller} {et~al.}(1998){Appenzeller}, {Fricke}, {F{\"u}rtig},
  {G{\"a}ssler}, {H{\"a}fner}, {Harke}, {Hess}, {Hummel}, {J{\"u}rgens},
  {Kudritzki}, {Mantel}, {Meisl}, {Muschielok}, {Nicklas}, {Rupprecht},
  {Seifert}, {Stahl}, {Szeifert}, \& {Tarantik}}]{FORS}
{Appenzeller}, I., {Fricke}, K., {F{\"u}rtig}, W., {et~al.} 1998, The
  Messenger, 94, 1

\bibitem[{{Butterley} {et~al.}(2020){Butterley}, {Sarazin}, {Le Louarn},
  {Osborn}, \& {Farley}}]{SCIDAR2}
{Butterley}, T., {Sarazin}, M., {Le Louarn}, M., {Osborn}, J., \& {Farley}, O.
  J.~D. 2020, in Society of Photo-Optical Instrumentation Engineers (SPIE)
  Conference Series, Vol. 11448, Adaptive Optics Systems VII, ed.
  L.~{Schreiber}, D.~{Schmidt}, \& E.~{Vernet}, 114481W

\bibitem[{{Cameron} {et~al.}(2009){Cameron}, {Britton}, \&
  {Kulkarni}}]{Cameron2009}
{Cameron}, P.~B., {Britton}, M.~C., \& {Kulkarni}, S.~R. 2009, \aj, 137, 83

\bibitem[{{Cl{\'e}net} {et~al.}(2015){Cl{\'e}net}, {Gendron}, {Gratadour},
  {Rousset}, \& {Vidal}}]{Clenet2015}
{Cl{\'e}net}, Y., {Gendron}, E., {Gratadour}, D., {Rousset}, G., \& {Vidal}, F.
  2015, \aap, 583, A102

\bibitem[{{Conan} {et~al.}(2003){Conan}, {Avila}, {S{\'a}nchez}, {Ziad},
  {Martin}, {Borgnino}, {Harris}, {Gonz{\'a}lez}, {Michel}, \&
  {Hiriart}}]{SanPedro}
{Conan}, R., {Avila}, R., {S{\'a}nchez}, L.~J., {et~al.} 2003, in Revista
  Mexicana de Astronomia y Astrofisica Conference Series, Vol.~19, Revista
  Mexicana de Astronomia y Astrofisica Conference Series, ed.
  I.~{Cruz-Gonzalez}, R.~{Avila}, \& M.~{Tapia}, 31--36

\bibitem[{{Dali Ali} {et~al.}(2010){Dali Ali}, {Ziad}, {Berdja}, {Maire},
  {Borgnino}, {Sarazin}, {Lombardi}, {Navarrete}, {Vazquez Ramio}, {Reyes},
  {Delgado}, {Fuensalida}, {Tokovinin}, \& {Bustos}}]{Ali}
{Dali Ali}, W., {Ziad}, A., {Berdja}, A., {et~al.} 2010, \aap, 524, A73

\bibitem[{{Irwin}(1985)}]{Irwin}
{Irwin}, M.~J. 1985, \mnras, 214, 575

\bibitem[{{Lazorenko}(2002)}]{Non_Kolm}
{Lazorenko}, P.~F. 2002, \aap, 396, 353

\bibitem[{{Lazorenko}(2006)}]{Lazorenko2006}
{Lazorenko}, P.~F. 2006, \aap, 449, 1271

\bibitem[{{Lazorenko} \& {Lazorenko}(2004)}]{LazLaz}
{Lazorenko}, P.~F. \& {Lazorenko}, G.~A. 2004, \aap, 427, 1127

\bibitem[{{Lazorenko} {et~al.}(2009){Lazorenko}, {Mayor}, {Dominik}, {Pepe},
  {Segransan}, \& {Udry}}]{Lazorenko2009}
{Lazorenko}, P.~F., {Mayor}, M., {Dominik}, M., {et~al.} 2009, \aap, 505, 903

\bibitem[{{Lazorenko} {et~al.}(2014){Lazorenko}, {Sahlmann}, {S{\'e}gransan},
  {Mart{\'{\i}}n}, {Mayor}, {Queloz}, \& {Udry}}]{PALTA2}
{Lazorenko}, P.~F., {Sahlmann}, J., {S{\'e}gransan}, D., {et~al.} 2014, \aap,
  565, A21

\bibitem[{{Le Louarn} {et~al.}(2000){Le Louarn}, {Hubin}, {Sarazin}, \&
  {Tokovinin}}]{MCAO}
{Le Louarn}, M., {Hubin}, N., {Sarazin}, M., \& {Tokovinin}, A. 2000, \mnras,
  317, 535

\bibitem[{{Lindegren} {et~al.}(2021){Lindegren}, {Klioner}, {Hern{\'a}ndez},
  {Bombrun}, {Ramos-Lerate}, {Steidelm{\"u}ller}, {Bastian}, {Biermann}, {de
  Torres}, {Gerlach}, {Geyer}, {Hilger}, {Hobbs}, {Lammers}, {McMillan},
  {Stephenson}, {Casta{\~n}eda}, {Davidson}, {Fabricius}, {Gracia-Abril},
  {Portell}, {Rowell}, {Teyssier}, {Torra}, {Bartolom{\'e}}, {Clotet},
  {Garralda}, {Gonz{\'a}lez-Vidal}, {Torra}, {Abbas}, {Altmann}, {Anglada
  Varela}, {Balaguer-N{\'u}{\~n}ez}, {Balog}, {Barache}, {Becciani}, {Bernet},
  {Bertone}, {Bianchi}, {Bouquillon}, {Brown}, {Bucciarelli}, {Busonero},
  {Butkevich}, {Buzzi}, {Cancelliere}, {Carlucci}, {Charlot}, {Cioni},
  {Crosta}, {Crowley}, {del Peloso}, {del Pozo}, {Drimmel}, {Esquej}, {Fienga},
  {Fraile}, {Gai}, {Garcia-Reinaldos}, {Guerra}, {Hambly}, {Hauser},
  {Jan{\ss}en}, {Jordan}, {Kostrzewa-Rutkowska}, {Lattanzi}, {Liao}, {Licata},
  {Lister}, {L{\"o}ffler}, {Marchant}, {Masip}, {Mignard}, {Mints}, {Molina},
  {Mora}, {Morbidelli}, {Murphy}, {Pagani}, {Panuzzo}, {Pe{\~n}alosa Esteller},
  {Poggio}, {Re Fiorentin}, {Riva}, {Sagrist{\`a} Sell{\'e}s}, {Sanchez
  Gimenez}, {Sarasso}, {Sciacca}, {Siddiqui}, {Smart}, {Souami}, {Spagna},
  {Steele}, {Taris}, {Utrilla}, {van Reeven}, \& {Vecchiato}}]{Lind2021}
{Lindegren}, L., {Klioner}, S.~A., {Hern{\'a}ndez}, J., {et~al.} 2021, \aap,
  649, A2

\bibitem[{{Masciadri} {et~al.}(2013){Masciadri}, {Lascaux}, \& {Fini}}]{MOSE}
{Masciadri}, E., {Lascaux}, F., \& {Fini}, L. 2013, \mnras, 436, 1968

\bibitem[{{Mendez} {et~al.}(2013){Mendez}, {Silva}, \& {Lobos}}]{cramer}
{Mendez}, R.~A., {Silva}, J.~F., \& {Lobos}, R. 2013, \pasp, 125, 580

\bibitem[{{Osborn} {et~al.}(2018){Osborn}, {Wilson}, {Sarazin}, {Butterley},
  {Chac{\'o}n}, {Derie}, {Farley}, {Haubois}, {Laidlaw}, {LeLouarn},
  {Masciadri}, {Milli}, {Navarrete}, \& {Townson}}]{Osborn2018}
{Osborn}, J., {Wilson}, R.~W., {Sarazin}, M., {et~al.} 2018, \mnras, 478, 825

\bibitem[{{Pott} {et~al.}(2018){Pott}, {Rodeghiero}, {Riechert}, {Massari},
  {Fabricius}, {Arcidiacono}, \& {Davies}}]{SPIE_ELT}
{Pott}, J.-U., {Rodeghiero}, G., {Riechert}, H., {et~al.} 2018, in Society of
  Photo-Optical Instrumentation Engineers (SPIE) Conference Series, Vol. 10702,
  Ground-based and Airborne Instrumentation for Astronomy VII, 1070290

\bibitem[{{Pravdo} \& {Shaklan}(1996)}]{Pravdo1996}
{Pravdo}, S.~H. \& {Shaklan}, S.~B. 1996, \apj, 465, 264

\bibitem[{{Rodeghiero} {et~al.}(2021){Rodeghiero}, {Arcidiacono}, {Pott},
  {Perera}, {Pariani}, {Magrin}, {Riechert}, {Gl{\"u}ck}, {Gendron}, {Massari},
  {Sauter}, {Fabricius}, {H{\"a}berle}, {Me{\ss}linger}, {Davies}, {Ciliegi},
  {Lombini}, \& {Schreiber}}]{ELT2}
{Rodeghiero}, G., {Arcidiacono}, C., {Pott}, J.-U., {et~al.} 2021, Journal of
  Astronomical Telescopes, Instruments, and Systems, 7, 035005

\bibitem[{{Sahlmann} {et~al.}(2015{\natexlab{a}}){Sahlmann}, {Burgasser},
  {Mart{\'{\i}}n}, {Lazorenko}, {Bardalez Gagliuffi}, {Mayor}, {S{\'e}gransan},
  {Queloz}, \& {Udry}}]{Sahlmann2015DE0823}
{Sahlmann}, J., {Burgasser}, A.~J., {Mart{\'{\i}}n}, E.~L., {et~al.}
  2015{\natexlab{a}}, \aap, 579, A61

\bibitem[{{Sahlmann} {et~al.}(2016){Sahlmann}, {Lazorenko}, {Bouy},
  {Mart{\'\i}n}, {Queloz}, {S{\'e}gransan}, \& {Zapatero Osorio}}]{GTC}
{Sahlmann}, J., {Lazorenko}, P.~F., {Bouy}, H., {et~al.} 2016, \mnras, 455, 357

\bibitem[{{Sahlmann} {et~al.}(2014){Sahlmann}, {Lazorenko}, {S{\'e}gransan},
  {Mart{\'{\i}}n}, {Mayor}, {Queloz}, \& {Udry}}]{PALTA1}
{Sahlmann}, J., {Lazorenko}, P.~F., {S{\'e}gransan}, D., {et~al.} 2014, \aap,
  565, A20

\bibitem[{{Sahlmann} {et~al.}(2015{\natexlab{b}}){Sahlmann}, {Lazorenko},
  {S{\'e}gransan}, {Mart{\'{\i}}n}, {Mayor}, {Queloz}, \&
  {Udry}}]{Sahlmann_palta3}
{Sahlmann}, J., {Lazorenko}, P.~F., {S{\'e}gransan}, D., {et~al.}
  2015{\natexlab{b}}, \aap, 577, A15

\bibitem[{Sarazin {et~al.}(2017)Sarazin, Osborn, Chacon-Oelckers, D{\'e}rie,
  Louarn, Milli, Navarrete, \& Wilson}]{SCIDAR1}
Sarazin, M.~S., Osborn, J., Chacon-Oelckers, A., {et~al.} 2017, in Optics in
  Atmospheric Propagation and Adaptive Systems XX, ed. K.~U. Stein \&
  S.~Gladysz, Vol. 10425, International Society for Optics and Photonics
  (SPIE), 104250B

\bibitem[{{Sch{\"o}ck} {et~al.}(2014){Sch{\"o}ck}, {Do}, {Ellerbroek},
  {Gilles}, {Herriot}, {Meyer}, {Suzuki}, {Wang}, \& {Yelda}}]{TMT}
{Sch{\"o}ck}, M., {Do}, T., {Ellerbroek}, B.~L., {et~al.} 2014, in Society of
  Photo-Optical Instrumentation Engineers (SPIE) Conference Series, Vol. 9148,
  Adaptive Optics Systems IV, ed. E.~{Marchetti}, L.~M. {Close}, \& J.-P.
  {Vran}, 91482L

\bibitem[{Taheri {et~al.}(2022)Taheri, McConnachie, Turri, Massari, Andersen,
  Bono, Fiorentino, Venn, VÃcran, \& Stetson}]{GeMS}
Taheri, M., McConnachie, A., Turri, P., {et~al.} 2022, The Astronomical
  Journal, 163, 187

\bibitem[{{Trippe} {et~al.}(2010){Trippe}, {Davies}, {Eisenhauer}, {Schreiber},
  {Fritz}, \& {Genzel}}]{Trippe}
{Trippe}, S., {Davies}, R., {Eisenhauer}, F., {et~al.} 2010, \mnras, 402, 1126

\end{thebibliography}

\newpage
\clearpage

\begin{appendix}

\section{Calibration of the model and observed data}{\label{caldata}}
\subsection{Rejection of  outliers}{\label{qq}}
{   In this study,  DIM  variance ${\sigma}_\mathrm{a}^2$ is found} analyzing the differences  $\langle \Delta ^2 \rangle - \sigma_\mathrm{p}^2  - \sigma_\mathrm{r}^2$. The estimated value of ${\sigma}_\mathrm{a}^2$ therefore can be affected by the  outlier measurements  in the analyzed sample of $\Delta$. The outlier  distribution besides  can depend  on $R$ producing a false increase of $\langle \Delta ^2 \rangle$ in a similar way affecting the  ${\sigma}_\mathrm{a}^2$  dependence on $R$. The first impression is that the influence of outliers is minor, but  really it is significant because the amplitude of ${\sigma}_\mathrm{a}^2$ is only 20--50\% of   $\langle \Delta ^2 \rangle$  for most  stars at  $R=2\arcmin$, decreasing even to  a few  percent only at  $R \sim 1\arcmin$. Thus even a minor bias in $\langle \Delta ^2 \rangle$  is amplified in the estimated value of ${\sigma}_\mathrm{a}^2$, especially at small $R$. Therefore we had to perform a special  calibration of the   residuals $\Delta$ to make its sample  robust to the outliers rejection  and then additionally  scaling  $\langle \Delta ^2 \rangle$ values to remove secondary  bias caused by the rejection of outliers. 

The outliers are removed usually at 3-sigma, or when $|\Delta|>3 \sigma_\mathrm{mod}$. This is well justified when  the residuals are distributed by a  Gaussian law. In our case,   residuals $\Delta$  are expected to follow this distribution with a sigma parameter $\sigma_\mathrm{mod}$ because the main error components $\sigma_\mathrm{p}$ and $\sigma_\mathrm{r}$ are caused by fluctuations of the photon number, which are  Gaussian by nature. To verify this,  we computed   normalized deviations  $u=\Delta/ \sigma_\mathrm{mod}$ and formed the frequency distribution of $u$. It was made separately for primary  and secondary P-stars, also, for each $R$, $k$ and for data accumulated   over all $R$. Surprisingly, accumulated distribution for all $\sim 2.5 \cdot 10^6$ residuals available  was found to be really very much like the Gaussian,  with no visual difference between the two P-star types or mode $k$. However, if the data are split  separately for the narrow and wide $R$ fields, the difference in distributions becomes essential. Thus, it reveals  an excess  at the distribution peak ($u \sim 0$) by 5--20\% for narrow $R$ and 5--10\% depletion for the measurements with wide $R$.  It means that the model uncertainties $\sigma_\mathrm{mod}$ are overestimated for narrow fields and underestimated for wide fields. This is due to insufficiently good error model, which makes the outlier rejection a little complicated.

\begin{figure}[tbh]
\includegraphics[width=\linewidth]{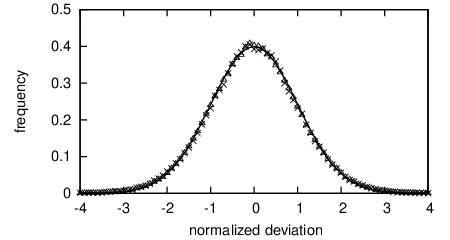}
\caption{Frequency distribution of the normalized positional residuals $u'$ obtained with the reduction mode $k=10$ in an example sky field Nr.19 for the measurements reduced separately for narrow  $R< 360$~px (triangles) and wide $R>840$~px (crosses) fields. The normal law distribution is shown with a solid line.}
\label{gist}
\end{figure}

We have found that any of the above partial distributions can be converted to being nearly Gaussian by introducing, instead of  $\sigma_\mathrm{mod}^2$, a new  variance $\sigma_\mathrm{mod}^2 /\alpha^2$, where $\alpha$ is a  constant (for some sky field and each $R$) scaling parameter, which makes the distribution of the deviations $u'=\alpha \Delta/ {\sigma}_\mathrm{mod}$  similar  to  the normal law distribution  (Fig.\,\ref{gist}). Values $\alpha$  were derived by minimizing deviation between the measured and the normal distribution and they were obtained for each sky field, $R$, $k$, and  separately for the target and reference stars. Most of values $\alpha$ are  in the range of 0.9--1.1. 

The sample distribution of  deviations $u'$ is little dependent on $R$, $k$, and the star type (primary or reference P-star) in any sky field.  This allowed us to perform rejection of outliers which exceed  $Z {\sigma}_\mathrm{mod} /\alpha$ at some cut level $Z$,  not introducing a bias dependent on $R$. Thus we saved best dependence of the DIM  variance on $R$. We applied  levels  $Z=2.5$, 3, 4, $Z= \infty$ (no rejection) and whereupon    abandoned the terms $\alpha$.  

The rejection procedure of course results in the decrease of the variance  $\langle \Delta ^2 \rangle $ depending on $Z$. For  a normal distribution this effect is easily estimated and can be corrected considering that after applying  rejection at level  $Z$  the variance is  $\gamma^2(Z)  =  \frac{1}{\sqrt{2\pi}}  \int_{-Z}^{Z}x^2 \exp(-x^2/2) dx <1$. Table\,\ref{var} provides  values $\gamma^2(Z)$ for normal distribution with cut levels $Z$ used in this investigation.  Thus, assuming normal law distribution of the residuals $\Delta/\sigma_\mathrm{mod} $ and having the variance $\langle \Delta ^2 \rangle_Z $ computed at some rejection level  $Z$, we can restore its unbiased variance
\begin{equation}
\label{eq:qq}
\langle \Delta ^2 \rangle_{\infty}= \langle \Delta ^2 \rangle_Z / \gamma^2(Z) 
\end{equation}
correspondent to the full data sample ($Z=\infty$). 

Similar numeric estimates of $\gamma^2(Z)$, showing the decrease of the variance, were computed for our measurements as the average of $({\Delta}/{\sigma_\mathrm{mod}})^2$ at each $Z$. We did it for all measurements of P-stars at all $R$, for each sky field, averaged  over  $M$ exposures,  and the resulting median for each $k$ after normalization to a unit value at $Z=4$ are given in Table\,\ref{var}. These estimates match well  those for the normal distribution mostly within $\pm 0.02$ and are little dependent on $k$. We consider this result acceptably good and as independent verification of the normal law distribution  of  single exposure residuals $\Delta$. This means also validity of the correction  (\ref{eq:qq}) for our data. Finally, we applied rejection with $Z=3$ only,   recomputed $\langle \Delta ^2 \rangle$ as the weighted average of single frame variances ${\Delta}^2$ and, using  Eq.\,\ref{eq:qq}, corrected these values with  $\gamma^2(3)$ dependent on $k$ (Table\,\ref{var}). 

\begin{table}[tbh]
\caption [] {Ratio  $\gamma^2$ for each reduction mode $k$ separately and for a normal distribution. }
\centering
\begin{tabular}{@{}c|cccc@{}}
\hline
\hline
$k$                                      &\multicolumn{4}{c}{  $Z$ }    \rule{0pt}{11pt}\\
\cline{2-5}
   & 2.5    & 3       & 4             & $\infty$    \rule{0pt}{11pt}\\
\hline
6  &  0.870 & 0.944 & 1.000  & 1.016 \rule{0pt}{11pt}\\
8  &  0.876 & 0.950 & 1.000  & 1.014\\
10  &  0.892 & 0.961& 1.000   & 1.014\\
12  &  0.889 & 0.959 & 1.000   & 1.012\\
14  &  0.895 & 0.965 & 1.000  & 1.011\\
\hline
normal distribution &  0.900 & 0.971  & 0.999 & 1.000  \rule{0pt}{11pt}\\
\hline
\end{tabular}
\label{var}
\end{table}

\subsection{Calibration of the error model components}{\label{s_cal}}
According to Sect.\,\ref{qq}, ${\Delta}/ {\sigma}_\mathrm{mod}$ for  single exposures  follow a normal law distribution. It allows us to estimate uncertainty of $\langle \Delta ^2 \rangle$, which is the weighted  average for a series of $M$ exposures. It is known that a sum of squares of  $n$  normal values has a chi-square  distribution with mathematical expectation $n$ and a variance $2n$. In our case, $n=M$ because $\Delta^2$ was defined as the average of this term in $x$ and $y$. Therefore,  mathematical expectation of   $\langle \Delta ^2 \rangle$  is equal to ${\sigma}_\mathrm{mod}^2$, and the uncertainty of this estimate is
\begin{equation}
\label{eq:sig_av}
\sigma_{\langle \Delta ^2 \rangle}= E {\sigma}_\mathrm{mod}^2 M^{-0.5},
\end{equation}
with $E=1$ for exactly normal distribution of the residuals $\Delta$. Ideally,  the differences $\langle \Delta ^2 \rangle - {\sigma}_\mathrm{mod}^2$ should have a zero expectation and   nearly a Gaussian distribution with sigma parameter $\sigma_{\langle \Delta ^2 \rangle}$. Due to a large number $M$ of images,  a bias  of  the order of 5\%  in the  model of ${\sigma}_\mathrm{mod}^2$  is already detected as excessive value of deviations $\langle \Delta ^2 \rangle - {\sigma}_\mathrm{mod}^2$. We searched for these inconsistencies  and removed them by calibrations described below. This step of the study is quite necessary because inaccuracy in the initial ${\sigma}_\mathrm{mod}^2$ model  were found to mount to a few $\sigma_{\langle \Delta ^2 \rangle}$.

\begin{figure}[tbh]
\includegraphics[width=\linewidth]{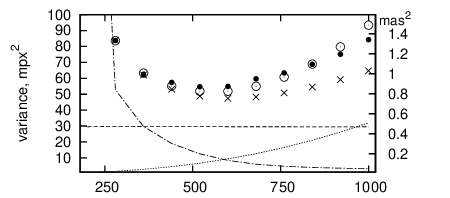}
\includegraphics[width=\linewidth]{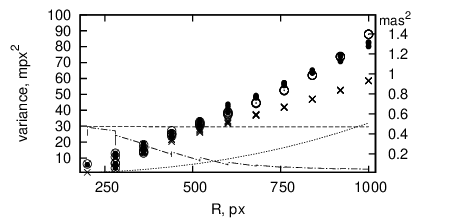}
\caption{Observed variance of the positional residuals $\langle \Delta ^2 \rangle$ (black dots) derived with an uncertainty of  about 1--2~mpx$^2$ as a function of the reference field radius for an example P-star in a sky field Nr.19. The initial model variance  $\sigma_\mathrm{mod}^2$ (crosses) and its components  $\sigma_\mathrm{p}^2$ (dashed line), $\sigma_\mathrm{r}^2$ (dash-dotted),  $\sigma_\mathrm{a}^2$ (dotted); updated (Sect.\ref{sectds})  model  variance  $\hat{\sigma}_\mathrm{mod}^2$ (circles).  The star was reduced as a primary  ({\it upper panel}) or a reference star ({\it lower panel}) using reduction mode $k=10$.   }
\label{demo}
\end{figure}

In each sky field, we have at hand $N_R=11$  estimates of the variances $\langle \Delta ^2 \rangle$ for each P-star reduced as a target, and yet $ (N_p-1)N_R  $ estimates for this star reduced as a reference object.  Fig.\,\ref{demo} illustrates typical change of the measured variance $\langle \Delta ^2 \rangle$ and its model components along the field size $R$ for an example P-star in a  sky field  Nr.19.   The upper panel corresponds to the primary status of this star and the error model Eq.\,\ref{eq:sig_oc0}.  Deviations between the measured and the model variances is visually seen and largely exceed the uncertainty $\sigma_{\langle \Delta ^2 \rangle}$, which is only about 2~mpx$^2$ at $R=1000$~px because of a large $M=2388$.

The lower panel corresponds to the measurements of positional residuals of the same star  $i$ but now treated as a reference  for  nearby  five primary P-stars,  which results in a multiple set of $\langle \Delta ^2 \rangle$ and ${\sigma}_{\mathrm{r}}$ values  shown in the plot.  Unlike the upper plot,  $\langle \Delta ^2 \rangle $  turns to zero at small $R$.   This is  a consequence of the fact that any errors in the photocenter  measurements of a bright reference star cause bias in the transformation fit functions which results in  equivalent change of the transformed positions  and so to small residuals $\Delta$.

\subsubsection{Calibration withing  sky fields}{\label{internal}}

Figure\,\ref{demo}  demonstrates that  model Eq.\,\ref{eq:sig_oc2b} represents the measured variance $\langle \Delta ^2 \rangle$ of the residuals far not perfect.  In particular, this can be due to use of   approximate parameters for  the DIM   taken from \citet{Lazorenko2006}. However,  direct fit of  a full  set of $\langle \Delta ^2 \rangle $ by function Eq.\,\ref{eq:sig_at}  via parameters $a$ and $b$ leads to a large discrepancy  between the measured and  updated model  variances. Besides,  we derived  strongly deviating estimates of $a$, $b$ found separately for different  sky fields, which is unacceptable. The problem therefore is in the approximate model values $\sigma_\mathrm{p}^2$ and $\sigma_\mathrm{r}^2$ which require correction at the level of $\sim $10--20\%. For instance, this may occur when the estimates (\ref{eq:sig_ph}) of $\sigma_\mathrm{p}^2$ are biased by errors in $I_G$ flux,  by deviations between the actual  and the model star image shape, blending, other effects (Sect.\ref{prec}). 

We assumed that the components $\sigma_\mathrm{p}^2$ and $\sigma_\mathrm{r}^2$  require  correction by a small scaling factor $\delta$ peculiar for each primary P-star. For instance,  the actual value of  $\sigma_\mathrm{p}^2$ for these stars $i$ is  $\hat{\sigma}_{\mathrm{p},i}^2 =(1+\delta_{\mathrm{p},i})\sigma_{\mathrm{p},i}^2$ and the  reference frame variance is $\hat{\sigma}_{\mathrm{r},i}^2= (1+\delta_{\mathrm{r},i}) \sigma_{\mathrm{r},i}^2 + g_i$ with the scaling factor $\delta_{\mathrm{r},i}$ related to $i$-th primary star, which is constant for all $R$, and  $g_i$ is an  offset. The DIM  component was updated to the model  $\hat{\sigma}_\mathrm{a}^2 = {a} R^{{b}}$ with new constants fit values ${a}$ and ${b}$  estimated for each sky field.   The calibration model contains  $3N_p + 2$  parameters ($\delta_{\mathrm{p},i}$, $\delta_{\mathrm{r},i}$, $g_i$, $a$ and $b$)  to be fit within each star field and which for primary stars have a simple structure:
\begin{equation}
\label{eq:m0}
\langle \Delta ^2 \rangle_i - \sigma_{\mathrm{p},i}^2 - \sigma_{\mathrm{r},i}^2  = \delta_{\mathrm{p},i} \sigma_{\mathrm{p},i}^2+ \delta_{\mathrm{r},i}  \sigma_{\mathrm{r},i}^2+ g_i + {a} R^{b}. 
\end{equation}
We note that corrections $\delta_{\mathrm{r},i}$ refer to the case when the star $i$ is  a primary probing object. In the case of  $i$-th star used as reference for $i'$ P-star, we first transformed the variance $\sigma_{\mathrm{r},i}^2$ (equal to $\sigma_{\mathrm{r},P_i>0}^2$ in Eq.\,\ref{eq:sig_pi}) to the variance $\sigma_{\mathrm{r},P_i=0}^2$ which emulate reference frame noise  related to the star $i'$ at  location of $i$-th star. Because  $i$ and $i'$ are nearby stars, they have nearly same set of reference stars and  the reference frame noise for these two stars with a status `primary' is approximately equal. Thus  $  \sigma_{\mathrm{r},P_i =0}^2$ can serve as emulation  $\bar{\sigma}_{\mathrm{r},i}^2$ of the reference noise $\sigma_{\mathrm{r},i}^2$ for $i$-th star in a  `primary' status. In terms of this variable,  the model variance is expressed as $ \sigma_\mathrm{mod}^2= \sigma_\mathrm{p}^2+\sigma_\mathrm{a}^2 - \bar{\sigma}_{\mathrm{r},i}^2 (\sigma_\mathrm{p}^2 + \sigma_\mathrm{a}^2) / (\sigma_\mathrm{p}^2 + \sigma_\mathrm{a}^2 + \bar{\sigma}_{\mathrm{r},i}^2)$.  Hence, we arrived at the calibration model:
\begin{equation}
\label{eq:m1}
\langle \Delta ^2 \rangle_i -  \sigma_{\mathrm{p},i}^2 + \sigma_{\mathrm{r},i}^2 = \delta_{\mathrm{p},i} \frac{\partial  \sigma_\mathrm{mod}^2}{  \partial \sigma_\mathrm{p}^2} \sigma_{\mathrm{p},i}^2+ \delta_{\mathrm{r},i}  \frac{\partial  \sigma_\mathrm{mod}^2}{ \partial  \bar{\sigma}_{r}^2}\bar{\sigma}_{\mathrm{r},i}^2+ g_i + {a} R^{b},
\end{equation}
where partial derivatives  are obtained in algebraic form  in terms of  $\bar{\sigma}_{\mathrm{r},i}^2$ used instead of  ${\sigma}_{\mathrm{r},i}^2$  for this star in a "primary" status. 

The system of equations (\ref{eq:m0}) and (\ref{eq:m1})  are solved combined using  100 -- 400 measurements of $\langle \Delta ^2 \rangle$ within each sky field separately. The derived corrections $\delta_{\mathrm{p},i}$ and $\delta_{\mathrm{r},i}$ are mostly within $\pm 0.2$, and a new updated model,
\begin{equation}
\label{eq:updated0}
 \hat{\sigma}_\mathrm{mod}^2= \hat{\sigma}_p^2 \pm  \hat{\sigma}_r^2  + {a}R^{{b}},  
\end{equation}
 produced a significantly better match to $\langle \Delta ^2 \rangle$ (Fig.\,\ref{demo}).  Thus, while the rms of the initial normalized residuals $\langle \Delta ^2 \rangle - {\sigma}_\mathrm{mod}^2$ was  about  5 -- 7 in terms of a factor $E$ in Eq.\,\ref{eq:sig_av}, for the updated model, this value decreased to the median $E=1.71$ for a full dataset, that is, for observations obtained at any seeing conditions. If the data was split into four  narrow  ranges of  FWHM, this value decreased  to $E=1.11$. This is quite natural  because corrections $\delta$ are better defined for stable observation conditions.  Derived terms $E$ provide  valuable information on the actual accuracy of  P-stars observations and the error model quality, therefore in the following we use these  term $E$ value. 
 
However even with updated model, the estimates of  ${a}$ (reduced to some standard exposure and to zenith)  and ${b} $ terms in Eq.\,\ref{eq:updated0} were found to be deviating between individual fields  over the tolerance limits. Therefore, instead of $aR^b$ we adopted   numeric (non-functional) estimate {   of  DIM variance $\hat{\sigma}^2_\mathrm{a}  = \langle \Delta ^2 \rangle  -  \hat{\sigma}_{\mathrm{p}}^2  \mp \hat{\sigma}_{\mathrm{r}}^2   $ equal to   a  sum of the fit function ${a}R^{{b}} $ and the residuals to the least-squares fit Eq.\,\ref{eq:m1}  for each P-star measurement. In this definition,  the  DIM component remains best  unbiased. }

\subsubsection{Comparison  between individual sky fields}{\label{sectds}}
We preliminary reduced DIM variances  $\hat{\sigma}^2_\mathrm{a}$ to  equal 30~s exposures  and to zenith with the normalizing factor  $(\sec{z})^{-\nu-p-1}(T/T_0)$ predicted by approximation Eq.\ref{eq:L4}. We  expected the reduced values  to be approximately equal for all sky fields, but actually,  for some  fields, they have shown systematic (irrespective of $k$)  offsets from the value $\langle \hat{\sigma}^2_\mathrm{a}  \rangle$  averaged over all fields. The difference $\hat{\sigma}^2_\mathrm{a} -\langle \hat{\sigma}^2_\mathrm{a} \rangle$ was found to fluctuate around some average  specific for a field $n$ with an amplitude proportional to $\langle \hat{\sigma}^2_\mathrm{a} \rangle$ so that the relative excesses $\theta_{n,k}= (\hat{\sigma}^2_\mathrm{a} -\langle \hat{\sigma}^2_\mathrm{a} \rangle)/\langle \hat{\sigma}^2_\mathrm{a}  \rangle$  depend on $k$  rather weak. Thus terms $\theta_{n,k}$ are peculiar for the field $n$ only.  After averaging $\theta_{n,k}$ over five reduction modes  $k$ applied, we come to the quantities $\theta_{n}= (\sum_{k} \theta_{n,k})/5$ which are typically  within $\pm 0.2$. We consider that non-zero values  $\theta_n$ are probably due to actual fluctuations of the turbulent strength of $C_n^2$  or the wind velocity $V$ in Eq.\ref{eq:ai} measured in a limited number of observations (see also Sect.\,\ref{on_int}). Irrespective of interpretation,  we applied corrections  $(1+\theta_n)^{-1}$  to the DIM  variance  as a scaling factor  to make {   final calibrated estimates ${\sigma}^2_\mathrm{a}$ } best  consistent between fields. 

\section{Some calibrations based on the metric integral}{\label{int22}}
A weak dependence of  terms  $\langle\mu_{k,R} \rangle$ on $d$ (Fig.\,\ref{fgint}) allows us to estimate  FORS2 metric integrals  at any  $d$ via easily computed  $I_\mathrm{c}(d)$  as 
\begin{equation}
\label{eq:Ii_Ic}
I_{\mathrm{F}}(d)=  \langle\mu_{k,R} \rangle I_\mathrm{c}(d)
\end{equation}

Above, we have found that   $\mu_{i}\approx \mathrm{const}$   for some $i$ and $k$ in a field given. For that reason values $I_{\mathrm{F},i}$  available for  the distance $d_0$ (we did computed integrals for  $d_0= 16$~km only, all $R$ and $k$) can be transformed to any other  distance as 
\begin{equation}
\label{eq:Ii_d}
I_{\mathrm{F},i} (d)=F(d) I_{\mathrm{F},i} (d_0)(d/d_0)^{\nu+p} 
\end{equation}
using the quantity
\begin{equation}
\label{eq:F}
  F(d)=I_\mathrm{c}^* (d)/ I_\mathrm{c}^*(d_0),
 \end{equation}
which displays deviations from the narrow angle power law,  when this term is  a unit exactly. Values $F(d)$ are fully defined by the  standard  CSD and are  important calibration terms used for correct  transformation of  $I_{\mathrm{F},i} (d_0)$ known at $d=d_0$ to any other distance. In particular,  transformation (\ref{eq:Ii_d}) is necessary for modeling the DIM variance based on the vertical profile of  turbulence  (Sect.\,\ref{on_int}). Because $F(d)$ is a function of $R$ and $k$, we have computed it for all varieties of $R$ and $k$ used in the study.
              
We extended the above analysis by considering  integral   $I$  properties as a function of   field size $R$ at $d$ fixed and find many similar features (Fig.\,\ref{tr}). Evidently, integrals  $I$   follow approximate  power law $I(R) \sim R^{\nu +p}$ and  exact dependence follows from  definition of $y$ variable in Eq.\,\ref{eq:6_7} as a product of $dR$. This  leads  to the  expression
\begin{equation}
\label{eq:I_r}
I_{\mathrm{c}} (d_0,R)= I_{\mathrm{c}} (d_0,R_0) F(d_0R/R_0) (R/R_0)^{\nu+p} 
\end{equation}
indicating minor deviation from dependence $I(R) \sim R^{3.67}$ due to the presence of a slightly modulating factor $F(d_0R/R_0)$ needed to convert $I$ between different $R$. Derived expression is valid for   CSD but, via $\mu$ terms, is applicable for FORS2 fields also.

\section{Von Karman model of the turbulence spectrum}{\label{von}}
Model of differential image motion (Sect.\,\ref{at} ) derived in \citep{LazLaz} is based on assumption that power spectral density of a phase follows a power law $q^{-3-p}$ of  the spatial frequency $q$, as expected for fully developed Kolmogorov turbulence with infinite outer scale  $L_\mathrm{out}=\infty$. In von Karman model, equivalent power law   $ (q^2+ L_\mathrm{out}^{-2})^{(-3-p)/2}$ \citep{Clenet2015} contains finite outer scale term $L_\mathrm{out}$. The difference between models is equivalent to applying  an extra filter function $\varphi(q) = (1+ q^2L_\mathrm{out}^{-2})^{(-3-p)/2}$. Consequently,  Eq.\ref{eq:G} for a spectral density $G(q)$ derived in \citep{LazLaz} by application  of several filters specific for differential measurements and assuming  a power law $q^{-3-p}$, remains valid  for any $L_\mathrm{out}$ by including  filter function $\varphi(q)$ in this expression. Function $\varphi(q)$ also should also be included in the integral expression Eq.\ref{eq:int}.

Typical values of $L_\mathrm{out}$ at altitudes of 16--20~km  are mostly within 10--100~m  as  estimated by  \citep{SanPedro, Ali}. This is comparable to the largest  size of phase fluctuations $L_Q$ (Eq.\ref{eq:Lcut}) removed well by filter $Q(q)$ (Sect.\,\ref{filt}) even for low reduction mode $k=6$, and removed yet efficient for   higher $k$. Thus the DIM variance ${\sigma}^2_\mathrm{a} $ is nearly  insensitive to fluctuations at long scales which are inhibited either by  filter $Q(q)$ or by actually long $L_\mathrm{out}$ if it exceeds $L_Q$. For that reason, we do not try to estimate  $L_\mathrm{out}$.

Considering a minor impact of the outer scale on ${\sigma}^2_\mathrm{a} $, by numeric simulation we derived the approximate expression
\begin{equation}
 \label{eq:outer}
I_{\mathrm{finite}\,  L} = I_{\mathrm{infinite} \, L} [1-0.06 \,R\, (8/k)\, (50/L_\mathrm{out})^2 ]
\end{equation} 
for conversion between metric integrals with finite and infinite scales, valid for  $25\,\mathrm{m} <L_\mathrm{out}<100 $~m,   $8\,\mathrm{m}< D< 40$~m; here, $R$ is expressed in arcminutes.

\section{Dependence of FORS2 DIM variance on the integral $I$ for Von Karman model}{\label{vonM}}
Computation  of coefficients $A_{\mathrm{F}}(k)$ (Table\,\ref{coef_A_k})  in Sect.\,\ref{on_int1} found in assumption of an infinite outer scale $L_\mathrm{out}$  were repeated for von Karman model of the turbulent spectrum assuming that $L_\mathrm{out}=50$~m. For this purpose we updated metric integrals with Eq.\,\ref{eq:outer} and ran computations same as described in Sect.\,\ref{on_int1}. We noticed small systematic change of  $A_{\mathrm{F}}(k)$ values well compensated by updating former  factor $\tau_0$ value from 0.84 to  0.89. With this modification, derived coefficients $A_{\mathrm{F}}(k)$ were found  to be all within 3-sigma of their values related to Kolmogorov turbulence. Thus, a change of a single parameter $\tau_0 $ allowed us to compensate well variations in metric integrals $I$ caused by von Karman model and keep unchanged  the differences between observations and our model for any $k$. We conclude that fit of the observed data with any turbulence model is equally good, at least, for  $L_\mathrm{out}$> 25~m. Very important that this conclusion refers also to any other coefficients derived in this study, thus their estimates for von Karman and Kolmogorov turbulence are expected to be equal if  turbulent scales are of a reasonable size.

\end{appendix}

\end{document}